\documentclass[aps,prb,twocolumn,superscriptaddress]{revtex4-1}

\usepackage{graphicx}
\usepackage{xcolor}
\usepackage{hyperref}
\usepackage{amsmath}
\usepackage{natbib}
\usepackage{url}
\usepackage{verbatim}

\newcommand{\soleil}{Synchrotron SOLEIL, L'Orme des Merisiers, Saint-Aubin, BP 48, F-91192 Gif-sur-Yvette, France}
\newcommand{\lsi}{Laboratoire des Solides Irradi\'es, \'Ecole Polytechnique, CNRS, CEA/DRF/IRAMIS, Institut Polytechnique de Paris, F-91128 Palaiseau, France}
\newcommand{\etsf}{European Theoretical Spectroscopy Facility (ETSF)}
\newcommand{\esrf}{ESRF, 71 avenue des Martyrs, 38000 Grenoble, France}
\newcommand{\crismat}{Laboratoire CRISMAT, CNRS UMR 6508, ENSICAEN, 14050 Caen, France}

\newcommand{\sorbonne}{Laboratoire de Chimie Physique, Mati\`ere et Rayonnement, Sorbonne Universit\'e, CNRS, F-75252 Paris, France}

\begin{document}

\title{Resonant inelastic x-ray scattering study of doping and temperature dependence of low-energy excitations in La$_{1-x}$Sr$_x$VO$_3$ thin films}

\author{Kari Ruotsalainen}
\email[]{kari.ruotsalainen@helmholtz-berlin.de}

\affiliation{\soleil}
\affiliation{\lsi}

\author{Matteo Gatti}
\affiliation{\lsi}
\affiliation{\etsf}
\affiliation{\soleil}

\author{James M. Ablett}
\affiliation{\soleil}

\author{Flora Yakhou-Harris}
\affiliation{\esrf}

\author{Jean-Pascal Rueff}
\affiliation{\soleil}
\affiliation{\sorbonne}

\author{Adrian David}
\affiliation{\crismat}

\author{Wilfrid Prellier}
\affiliation{\crismat}

\author{Alessandro Nicolaou}
\email[]{alessandro.nicolaou@synchrotron-soleil.fr}
\affiliation{\soleil}

\date{\today}

\begin{abstract}
We present a temperature and doping dependent resonant inelastic X-ray scattering experiment at the V L$_{2,3}$ and 
O K edges in La$_{1-x}$Sr$_x$VO$_3$ thin films with $x$=0 and $x$=0.1. This material is a canonical example of a compound 
that exhibits a filling control metal-insulator transition and undergoes orbital ordering and antiferromagnetic transitions 
at low temperature.  Temperature dependent measurements at the V L$_{3}$ edge reveal an intra-t$_{2g}$ excitation that blueshifts 
by 40 meV from room temperature to 30 K at a rate that differs between the para- and antiferromagnetic phases. The lineshape can be 
partially explained by a purely local model using crystal field theory calculations. In the lightly doped regime, at $x=0.1$, the 
doping is shown to affect the local electronic structure primarily on the O sites, which is in disagreement with a simple Mott-Hubbard 
picture. Furthermore, we reveal the presence of phonon overtone features at the O K edge, which evidences that the low energy part of the 
spectrum is dominated by phonon response.

\end{abstract}


\maketitle

\section{Introduction}
The AVO$_3$ compounds  (A is a rare earth or Y) exhibit antiferromagnetism (AFM) and orbital ordering (OO) at low  
temperatures.\cite{Sagephd,MIYASAKA2003} In a local picture, the V atoms have a $d^2$ configuration in a slightly 
compressed octahedral local environment. Physical understanding of the AFM/OO  in AVO$_3$ relies on two views: 
the ordering transitions are associated with the Jahn-Teller effect or orbital superexchange between nearly 
degenerate nearest neighbor vanadium $t_{2g}$ electronic states.\cite{Mizokawa1996,Sawada1996,Mizokawa1999,Khaliullin2001,Motome2003,REN2003,Ulrich2003,Fang2004,Miyasaka2005,SOLOVYEV2006,DeRaychaudhury2007,Horsch2008,ZHOU2008,Rosciszewski2018} The splitting of the $t_{2g}$ states as a function of temperature
and doping is an important parameter for elucidating the physical drivers of the phase transitions.
The strength of the orbital superexchange is dependent on the V-O-V bond angles connecting adjacent 
octahedra. For bulk YVO$_3$ (YVO) the V-O-V angles remain below 145 degrees between room temperature and 
65 K, and it was concluded that crystal field splitting is the dominating factor.\cite{BLAKE2002,BENCKISER2013,Reul2012} 
Bulk LaVO$_3$ (LVO) exhibits V-O-V angles of within a range of 156 to 158 degrees in the same temperature range.\cite{BORDET1993}  
Strained epitaxial LVO is a particularly interesting  case\cite{Sclauzero2015,Sclauzero2016,He2012,Weng2010,Hotta2007,Jana2018} for 
further study as the structure is modified by the epitaxial growth constraints and the V-O-V angles straighten to 164-168 degrees.\cite{ROTELLA2015} 

Bulk LVO is a paramagnetic (PM) insulator at room temperature\cite{Inaba1995,MIYASAKA2000}. The AFM and OO transitions are found near $T = 140$ K in single crystals. \cite{Nguyen1995,MIYASAKA2000,MIYASAKA2003,ZHOU2008,TUNG2008}  
The room temperature crystal structure of LVO is orthorhombic ($Pnma$) and it distorts to a monoclinic 
structure ($P2_1/a$) upon entering the AFM/OO state.\cite{BORDET1993,REN2003,Khan2004} The structure of epitaxial LVO 
films grown on the (001) surface of SrTiO$_3$ (STO) has been studied with diffraction methods finding that structure 
can be refined assuming either $Pnma$ or $P2_1/m$ space groups. \cite{ROTELLA2012,ROTELLA2015,Hotta2006} Temperature 
dependent diffraction studies give no indication of a structural phase transition breaking the lattice symmetry.\cite{LEKSHMI2005,MELEY2018} 
The AFM transition in epitaxial films is however seen in temperature dependent magnetization measurements on LVO and in the related 
compound PrVO$_3$. The occurrence of the OO transition in LVO on STO has been deduced from Raman spectroscopy.
\cite{WANG2015,VREJOIU2016,VREJOIU2017} Recent investigations on strained LVO films by low-energy muon spin spectroscopy and 
ellipsometry measurements show that the compressive strain only reduces by few K the OO and AFM transition temperatures 
and do not suppress the bulk G-type orbital order. \cite{Meley2021} 

La$_{1-x}$Sr${_x}$VO$_3$ (LSVO) is considered a paradigmatic compound exhibiting a filling-control metal-insulator transition.\cite{Mott1990,Imada1998,FUJIMORI2001}
In LVO a 1.0$\pm$0.2 eV band gap opens between V $t_{2g}$ states, while V $e_g$ states are completely empty.\cite{Maiti2000} 
Within the Zaanen-Sawatzky-Allen classification,\cite{Zaanen1985} LVO belongs to the Mott-Hubbard regime.\cite{FUJIMORI2001} 
Sr doping induces a metal insulator transition in La$_{1-x}$Sr${_x}$VO$_3$ at $x$=0.17 and also reduces the AFM and OO critical 
temperatures and ultimately suppresses the transitions at $x=0.27$. \cite{Mahajan1992,Inaba1995,Imada1998,MIYASAKA2000,Fujioka2006}
In a pure Mott-Hubbard picture, Sr substitution leads to hole doping, converting V$^{3+}$ ions into V$^{4+}$. 
Core-level photoemission experiments on La$_{1-x}$Sr$_x$VO$_3$ have shown that the fraction of V$^{4+}$ 
with respect to V$^{3+}$ is indeed enhanced upon hole doping at a high Sr concentration of $x$=0.4.\cite{LEKSHMI2005} 
Since the Sr$^{2+}$ ions have larger radii than La$^{3+}$, the structure (un)distorts towards the cubic structure of SrVO$_3$ 
and the bandwidth also increases.\cite{Imada1998,FUJIMORI2001} Sr doping is expected to form an empty band of acceptor 
states near the Fermi level, which is progressively broadened with increasing doping until it overlaps with the 
filled $t_{2g}$ band giving rise to a metal.\cite{Egdell1984}

The electronic low-energy excitation spectrum of LVO has been characterized with optical spectroscopies,\cite{Inaba1995,Miyasaka2002,Fang2003,Fujioka2006,Kim2018} in which the crystal 
field transitions are weak due to being forbidden by the dipole selection rules. Resonant inelastic 
X-ray scattering (RIXS), on the other hand, excels at revealing crystal field transitions at 
L absorption thresholds of $3d$ transition metal atoms.\cite{AMENT2011} Recent advances\cite{BROOKES2018} 
in energy resolution have made detailed studies of low energy crystal field splittings even well below 
100 meV possible.\cite{AMORESE2018,AMORESE2019} Low energy excitations were clearly resolved in YVO 
employing RIXS at $\approx$ 60 meV resolution.\cite{BENCKISER2013}  This experiment focused on the 
dispersion of the low energy excitations across the phase diagram. Bulk LVO\cite{CHEN2015} and 
NdVO$_3$\cite{LAVEROCK2014} have been studied with RIXS at 0.4 eV resolution. Due to the modest 
energy resolution the lineshapes of low energy excitations could not be resolved. 

In this article we present a RIXS study of the temperature and doping dependencies of 
low-energy excitations in La$_{1-x}$Sr${_x}$VO$_3$ thin films with $x$=0 and $x$=0.1 
(LSVO). We analyze the effects of doping by comparing the V L and O K edge absorption and 
RIXS spectra recorded at moderate energy resolution. We find that the light Sr doping does 
not cause clear signatures of V$^{4+}$ ions, but rather affects the local electronic 
structure on the O site. Furthermore, using measurements with a state-of-the-art energy 
resolution, we show further that the intra-$t_{2g}$ excitations in LVO  blueshift by 40 
meV when cooling from room temperature to 30 K, and that the rate of the blueshift changes 
between the PM and AFM phase.    

\section{Experimental methods}

\begin{figure}
\includegraphics[width=1\columnwidth]{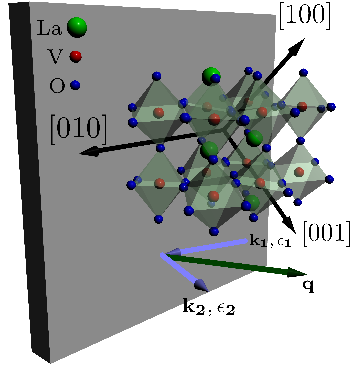}
\caption{\label{fig:expgeom} The crystal structure of strained LVO drawn from a refined P21/m structure\cite{ROTELLA2015} together with the experimental scattering geometry. Incident light defined by the photon energy $\hbar\omega_1$, wave vector $\vec{k}_1$ and polarization vector $\vec{\epsilon}_1$ is scattered to a state defined by $\hbar\omega_2$, $\vec{k}_2$ and $\vec{\epsilon}_2$, transferring energy $\hbar\omega=\hbar\omega_1-\hbar\omega_2$ and momentum $\hbar\vec{q}=\hbar\textbf{k}_1-\hbar\textbf{k}_2$. 
$\textbf{k}_1$ makes an angle of 20 degrees with the sample surface plane and the scattering angle is 85 degrees.}
\end{figure}

LVO and LSVO thin films  were grown on STO (001) substrates using the pulsed laser deposition (PLD) method. 
A KrF excimer laser ($\lambda$ = 248 nm) with a repetition rate of 2 Hz and laser fluence of $\approx$2 J/cm2 was 
focused on stoichiometric ceramic targets. All the films used in this study were deposited at an optimum 
growth temperature of 920 K and under oxygen partial pressure of 10$^{-6}$ mbar. PLD yields high quality 
epitaxial films of LVO as shown by structural, transport and optical studies.\cite{ROTELLA2012,ROTELLA2015,WANG2015}
We show the $P2_1/m$ structure of LVO on STO alongside with the scattering geometry in 
Fig. \ref{fig:expgeom}. The substrate plane normal is along the [101] direction of the 
LVO crystal structure and the film has a domain structure characterized by fourfold rotations about the plane normal.\cite{ROTELLA2012}
In terms of the V-O bonds, this means that long and short V-O bonds alternate along the surface of the sample. 
We performed the XAS measurements using circular left (CL) polarized light. The XAS and the RIXS spectra 
discussed later are considered averages over the appropriate incident and outgoing polarization states.

\begin{figure*}
\centering
\includegraphics[width=1\textwidth]{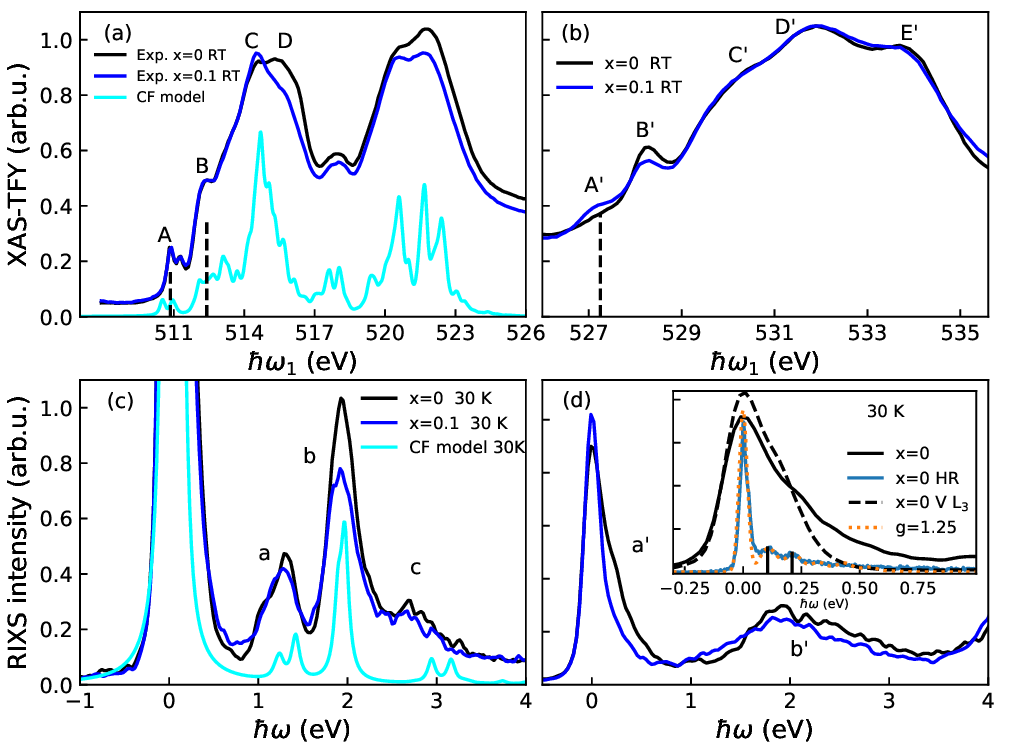}
\caption{ \label{fig:holedoping}   (a) Comparison of XAS-TFY spectra for LVO
and LSVO measured at room temperature (RT) using CL polarized light at the V L edges and a crystal field model (CF). Constant backgrounds were subtracted followed by normalization to the first pre-peak of the V L$_3$ edges (feature A). 
(b) The O K edge region. The spectra are normalized to D'.  
(c) Low temperature RIXS spectra recorded at feature B of Fig. \ref{fig:holedoping} (a). The spectra were processed by subtracting constant backgrounds and normalizing to the maximum of the feature b of the $x$=0 spectrum. The CF model calculation is scaled for presentability. 
(d) Low temperature O K edge RIXS spectra excited at feature A' of  panel (b). In the inset we compare low energy excitations observed at the V L$_3$ edge with the ones observed at the pre-edge region of O K edge. High (low resolution) RIXS spectra showing phonon excitations are compared against simulation within the Ament model\cite{Ament2011b} using a phonon energy of 105 meV and a coupling constant g=2.3(1.25).}.

\end{figure*}
RIXS experiments were performed both at V L$_{2,3}$ and O K edges using CL and linear-vertical (LV) polarized light. 
Several sets of measurements on different samples having the same doping were performed at the SEXTANTS 
beamline\cite{SACCHI2013} of the Synchrotron Soleil using the AERHA spectrometer.\cite{CHIUZBUAIAN2014} These measurements, 
performed with an overall energy resolution ranging from 100 meV to 180 meV (FWHM), were used to check the 
reproducibility of the experimental results, and to discover the temperature effect on the low energy 
losses measured at the V L$_3$ edge. Once the reproducibility and the robustness of the results were 
proven, high resolution measurements were carried out at the  ID32 beamline of the European Synchrotron 
Radiation Facility using the ERIXS spectrometer.\cite{BROOKES2018} The overall experimental resolution 
of approximately 40 meV allowed us to track precisely the temperature behavior of the energy loss peaks of interest. 
The X-ray absorption (XAS) spectra presented in this article were recorded during the experiments at SOLEIL 
using total fluorescence yield (TFY) detection. The probing depths of TFY and RIXS are of the order 100 nm. 
The former is more sensitive to the sample surface region as low energy fluorescence (e.g. V M shell, La N shell) is also detected. 

\section{Results and discussion}

\subsection{Hole doping of La$_{1-x}$Sr$_x$VO$_3$: Comparing V L and O K edge XAS and RIXS}

In this section we discuss the evolution of the electronic structure upon doping by comparing the XAS and RIXS measurements
for LVO and LSVO. XAS at the V L$_{2,3}$ edge is shown in Fig. \ref{fig:holedoping} (a) and the O K edge region is presented 
in (b). In (c) we present RIXS data acquired at the V L$_3$ edge exciting on B (512.4 eV). Finally in (d) 
we show RIXS spectra obtained at the O K edge by exciting on A$'$ (527.3 eV), where in the intermediate state 
an O 1s core electron is promoted to a hybridized O 2p/V $t_{2g}$ state.
Comparing the XAS measurements in Fig. \ref{fig:holedoping} (a) and (b), we note that while the lineshape of the 
low-energy part of the V L$_3$ edge (A, B) remains unchanged upon Sr to La substitution, the pre-edge region (A'and B') 
of the O K edge is strongly modified.

To understand how the hole doping affects the spectra and the electronic structure of LVO, 
we first analyze the character of the main features in the XAS spectra and we describe what is 
expected from the appearance of a d$^1$ contribution at the V L$_3$ edge.
The V L$_3$ XAS in a d$^2$ configuration and a quasi-O$_h$ crystal field is characterized by a 
double pre-peak structure with maxima located at 510.8 and 511.2 eV (A), a shoulder at 512.4 eV (B) 
and a main peak at 514.6 eV (C). All these features have been found in the XAS spectra of several 
vanadates including bulk LVO \cite{CHEN2015}, hybrid-MBE\cite{ZHANG2017} and pulsed laser 
deposition\cite{WADATI2009} grown films, bulk YVO\cite{PEN1999,BENCKISER2013} and bulk NdVO$_3$.\cite{LAVEROCK2014} 
The lineshapes of AVO$_3$ are essentially the same as in V$_2$O$_3$.\cite{PARK2000}  
Instead, in a 3d$^1$ configuration the double pre-peak feature A is absent\cite{schmitt2002} and the 
lineshape is dominated by a shoulder and a main peak that are found at higher energies.\cite{schmitt2002,HAVERKORT2005,PEN1999,PARK2000} 
The shoulder feature of d$^1$ overlaps with the high energy peak of A and hence the intensity ratio of the 
peaks at A should change by the introduction of d$^1$. As noted in the first paragraph, the spectra overlay from A to almost up to C.  
Therefore, the lineshape of the V L$_3$ XAS gives us no indication of d$^1$. 
This interpretation will receive further support from the RIXS results discussed lateron. 

Above C, the XAS spectra of LVO and LSVO start to deviate. We attribute this discrepancy to surface contamination. 
Photoemission experiments have indeed shown that the near surface layers of upcapped LVO films contain V$^{4+}$, and that their concentration is enhanced over time.\cite{WADATIthesis} Furthermore, we note that XAS measurements on YVO report a C/D ratio that evolves with temperature, but the authors did not emphasize this behavior.\cite{BENCKISER2013} We can therefore conclude that the XAS spectra representative of the bulk electronic structure of LVO and LSVO films remain unchanged at the V L$_3$ edge upon light hole doping, at odds with our observations at the O K edge. 

In order to compare O K-edge XAS spectra of LVO and LSVO we have normalized them by imposing the same intensity at D$'$. 
We used a different normalization factor with respect to the one used for the V L$_{23}$ edges in order to highlight the 
relative variations occurring in the pre-edge region (A$'$,B$'$). The evolution of the O K edge spectra as a function 
of hole doping would indeed be less visible by using a unique normalization factor for V and O edges. 
The deviation of the XAS spectra at V L$_{23}$ edges for energies above C is related to a different background 
originated by the different spectral weight present at D and affecting also the baseline before the O K edge. 
Therefore, our interpretation only relies on the differences observed in the XAS lineshapes. We do not compare O 
with respect to V intensities nor draw any conclusion based on the absolute intensities of the XAS features.

The O K-edge XAS spectra are characterized by two pre-edge peaks A$'$ and B$'$ at 527.3 eV and 528.2 eV, 
followed by structures C$'$ at 530.3 eV, D$'$ at 531.9 eV and E$'$ at 533.8 eV. 
The pre-peaks A$'$ and B$'$ have been assigned to exchange-split V t$_ {2g}$ states hybridized with 
O $2p$ and the peaks C$'$ and D$'$ to the e$_g$ states hybridized with O $2p$.\cite{MOSSANEK2009} 
Local density approximation (LDA) + Hubbard U calculations have also been used to assign features 
A$'$ and B$'$ similarly, C$'$ to O $2p$ hybridized with e$_g$ states, and D$'$ and E$'$ to La $5d$ 
states hybridized with O 2p.\cite{CHEN2015} Isoelectronic Lu doping was experimentally studied, and 
the experimental XAS results show no significant changes in intensity at the O K pre-edges even for 
pristine LuVO$_3$.\cite{CHEN2015} Therefore, the changes observed at A$'$ and B$'$ reveal the introduction 
of holes in the valence band. Hence, at least up to 10\% Sr replacement the doping does not seem to 
affect the V site electronic structure by a change in valency, but rather indirectly via the O orbitals 
that are hybridized with the t$_{2g}$ states of vanadium. Our results thus reveal a deviation from 
pure Mott-Hubbard like behavior. This conclusion is consistent with interpretation of photoemission and 
electron energy loss data on LSVO.\cite{Egdell1984} Similar behavior at light doping is also found in Y$_{1-x}$Ca$_x$VO$_3$.\cite{PEN1999} Cuprates and NiO, i.e. late transition metal oxides belonging to 
the different class of charge transfer insulators, also behave similarly.\cite{PhysRevLett.66.104,PhysRevB.45.1612}  
This shows how the classification to Mott-Hubbard and charge transfer systems represents an oversimplified picture 
for such a complex correlated oxide systems.

The absence of V d$^1$ states upon light doping is further confirmed by our RIXS measurements presented in Fig. \ref{fig:holedoping} (c). The crystal field excitations are labeled \textit{a-c}.  
Peak \textit{a} at 1.2 eV is a $\Delta S = 1$ intra-t$_{2g}$ transition with the low  (high) energy edge representing final states with 1 (2) electrons occupying the upper t$_{2g}$ states. 
Peak \textit{b} at 2 eV corresponds to S=1 final states with an electron transferred across the e$_g$-t$_{2g}$ crystal field gap. 
Excitations \textit{a} and \textit{b} reflect the Hund's exchange and the octahedral component of the crystal splitting between 
the t$_{2g}$ and $e_g$ states, respectively\cite{YUSHANKAI2013,DEGROOT2008}. Feature \textit{c} at 2.7 eV consists of higher 
energy terms with an excitation across the crystal field gap, and a high energy $\Delta S =1$ intra-t$_{2g}$ transition. 
These assignments were made using crystal field calculations (Sec. \ref{ssec:cft}) and are in agreement with p
revious  analyses.\cite{CHEN2015,LAVEROCK2014,BENCKISER2013} They are also consistent with a quantum chemistry calculation 
for VOCl (d$^2$).\cite{BOGDANOV2011}
Considering that peak \textit{a} is absent for a d$^1$ system (see e.g. LaTiO$_3$\cite{ULRICH2008}), 
if some of the holes are localized on V sites, the intensity of peak \textit{a} can only reduce. 
For peak \textit{b} one expects the intensity to enhance due to the d$^1$ component. The ratio 
of peak \textit{a} area to peak \textit{b} area is hence expected to be reduced. In our data the 
ratio is in fact enhanced excluding the presence of V d$^1$ sites. This can be understood by considering the 
influence of the rare-earth atom on the V-O hybridization. Chemical bonding in 3$d^1$ rare earth vanadates 
and titanates has been studied with LDA + dynamical mean field theory (DMFT) and show that the rare-earth 
atom states form $\sigma$ bonds with the same oxygen orbital that forms $\pi$ bonds with the 
metal t$_{2g}$.\cite{PAVARINI2005} Furthermore, the Sr contribution to the t$_{2g}$ states is 
lesser in SrVO$_3$ than the La contribution in LaVO$_3$.\cite{HAULE2014} The final states 
corresponding to peak \textit{b} have an occupied $e_g$ orbital and hybridize stronger with 
the neighboring O atoms than t$_{2g}$ type states. According to LDA+U calculations\cite{CHEN2015} 
the unoccupied e$_g$ states overlap in energy with the La 5d states. Therefore the intensity 
reduction seen particularly for peak \textit{a} and to some extent \textit{b} are explained the 
evolution of the hybridization in the RIXS final states. 
\subsection{Phonon and electronic excitations \\ at the O K-edge}{\label{ssec3b}}
In this section we discuss in details the low energy excitation spectrum measured at the O K edge. 
Low-energy excitations are observed in an overlapping energy range below 1 eV at both the V and O edges when the 
excitation energy is tuned over states with t$_{2g}$ character [ see the inset of Fig. \ref{fig:holedoping} (d)]. 
In order to discuss the origin of these excitations, we track their evolution with doping and we employ the 
state-of-the-art of energy resolution to resolve their components. Our results point toward a phonon origin 
of the low-energy excitations showing up at the O K edge, hence being of different nature with respect to 
the ones observed at the V L$_3$ edge, which have an orbital origin.

In Fig. \ref{fig:holedoping} (d) we compare RIXS spectra of LVO and LSVO. They were acquired at 30 K with 
an excitation energy of 527.3 eV [A$'$ in Fig. \ref{fig:holedoping} (b)]. In LVO we observe a shoulder near 
the elastic line, labeled \textit{a$'$}. This feature, which extends up to 0.7 eV, is strongly reduced in LSVO. 
This  means that such an excitation is dramatically affected by Sr to La substitution. This observation marks 
a clear difference with respect low-energy excitations we observed at the V L$_3$ edge, which appear very 
similar in LVO and LSVO [Fig. \ref{fig:result_tdep} (a)-(c)]. Another important dissimilarity is revealed by 
the high resolution data of LVO presented in the inset of Fig. \ref{fig:holedoping} (d). They allow us to 
clearly resolve peaks at 105 meV and 210 meV at the O K edge. Their constant energy spacing together with 
the small FWHM point to a phonon origin. Orbital excitation continua are indeed expected to be much broader 
in energy, consistently to our observations at the V L$_3$ edge. The high resolution spectrum, which was 
acquired by using an excitation energy detuned by +500 meV with respect with respect to A$'$ can be well described 
by using the eq. (3)\cite{Ament2011b} of the Ament model. The best agreement is reached by using an electron-phonon 
(el-ph) coupling constant g=2.3 and a phonon energy of 105 meV. For the low resolution spectrum, acquired at A$'$, 
only the low-energy part can be described by such a simple model. Furthermore, a lower g value of 1.25 must be 
used in order to reach a satisfactory agreement with the data.  
We think that, while the phonon origin of such a low-energy excitation showing up at the O K edge is certain, 
the extraction of a reliable el-ph coupling constant is not straightforward due to the complex process through 
which phonons are excitated during the RIXS process. The lack of overall agreement with the Ament model can be 
explained considering phonons and core hole screening. The neglect of mode coupling\cite{GEONDZHIAN2018,GEONDZHIAN2020} 
is probably a significant reason, as perovskite oxides have several close lying high energy optical branches.\cite{CHOUDHURY2008} 
The distortion of the intermediate state potential energy surface also affects the phonon intensities.\cite{GEONDZHIAN2020} 
That said, a full understanding of the phonon spectrum of LVO, including the identification of the phonon modes involved 
together with an precise determination of the el-ph coupling constant would require a dedicated study including several 
excitation energies and experimental geometries, which is beyond the scope of the present work.

In a previous RIXS study on YVO\cite{BENCKISER2013} the low energy excitations at the O K edge were interpreted as single and 
double orbital excitations. Our data point toward a different scenario. The clear presence of a phonon 
progression 
suggest that \textit{a$'$} exhibits dominant phonon contributions. The long tail towards higher energy 
losses would be related to multiple phonon excitations in response to core excitation.\cite{VazdaCruz2019} 
Moreover, our interpretation accounts for the strong suppression of a' by light doping as well, which would be 
due to disorder induced reduction of phonon lifetimes with the consequent broadening. 
This is consistent with Raman scattering results for Ca doping of YVO, where 2\% Ca replacement causes a 
significant broadening of the phonon excitations.\cite{fujioka2008}

It is interesting to remark that O K edge RIXS in molecular liquids features a long extending continuum of excitations 
when excited at O K pre-edge resonances.\cite{SUN2011,VazdaCruz2019,GEONDZHIAN2018} Hence a dielectric environment does 
not suppress the vibrational RIXS signal with respect to the gas phase.\cite{HENNIES2010} Moreover, several well resolved 
vibration harmonics were observed by RIXS at the N K edge in the \textit{k}-(BEDT-TTF)$_2$Cu$_2$(CN)$_3$ charge transfer salt.\cite{PhysRevB.96.184303} Considering long-range ordered oxides with 3d or lighter metal atoms, RIXS on the O K 
edge has been studied for e.g. Li$_2$CuO$_2$ and GeCuO$_3$\cite{MONNEY2013}, Al$_2$Si$_2$O$_5$(OH)$_4$\cite{ERTAN2017} and SrTiO$_3$\cite{DENIZ} i.e. in systems very different by their electronic structure and magnetic properties, yet 
continuum features extending past 0.5 eV energy losses are found. 

The O K edge RIXS presented in Fig. \ref{fig:holedoping} (d) reveals an additional feature labelled b', which is broad 
and centered at about 2 eV energy loss. This feature, which is also present in YVO, was attributed to excitations between 
the lower and upper Hubbard bands.\cite{BENCKISER2013} It was also noted that the onset energy of \textit{b$'$} in YVO 
matches rather well with the onset of optical conductivity for YVO.\cite{BENCKISER2013} Consistently to our data, showing 
the presence of more spectral weight at lower energy loss in LSVO, the onset of the optical conductivity of LSVO is 
found well below the one of LVO.\cite{fujioka2008} Doping induced disorder effects on the density of states of AVO$_3$ 
compounds in the vicinity of the band gap have been studied using a multiband Hubbard model 
Hamiltonian approach.\cite{AVELLA2018} They show defect induced unoccupied states appearing just below the upper 
Hubbard band, which is  likely at the origin of the evolution presented here.

\subsection{Temperature dependence of RIXS at the V L$_3$ pre-edge}{\label{sec_tdep}}

\begin{figure*}
\includegraphics[width=1\textwidth]{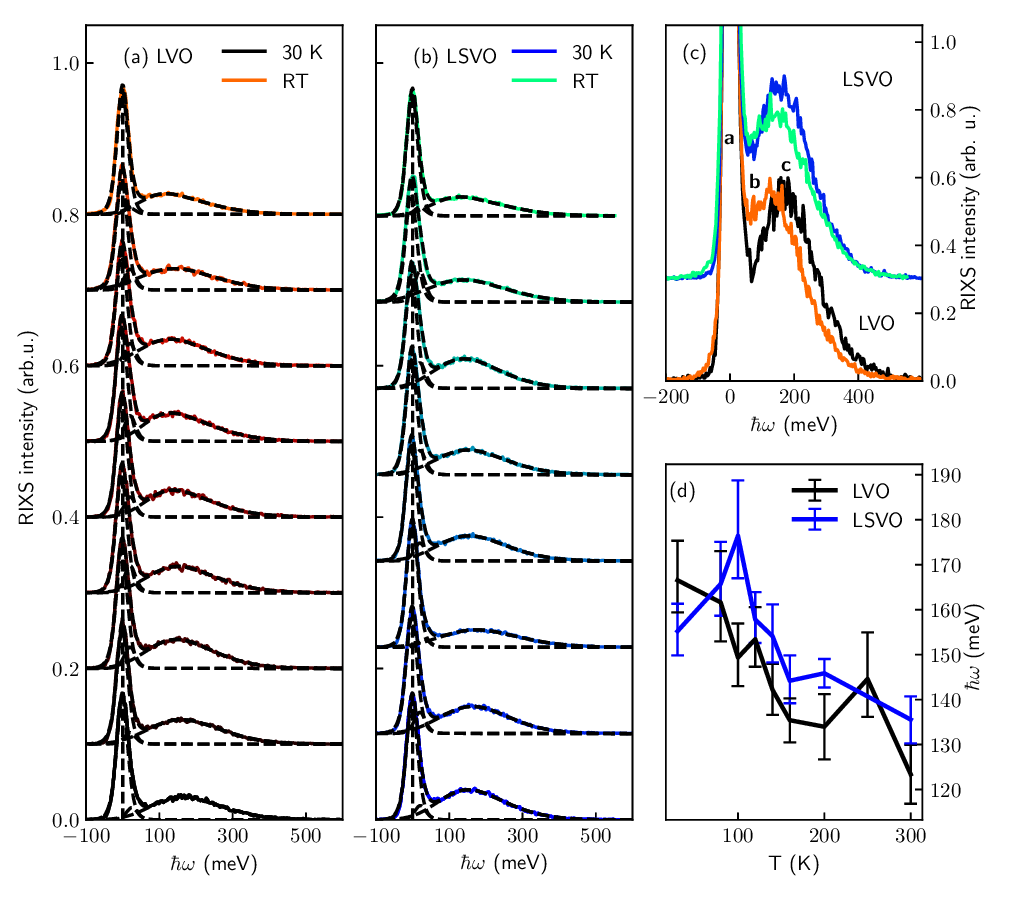}
\caption{\label{fig:result_tdep} Temperature dependent high resolution RIXS with 40 meV overall resolution. The presented spectra were measured with the incident photon energy tuned to feature A of Fig. \ref{fig:holedoping} (a).  
 (a) The temperature dependent RIXS spectra for $x=0$. The spectra are normalized to their maxima and they are ordered by increasing temperature. (b) The temperature series for $x=0.1$. (c) The endpoint temperature RIXS spectra for $x=0$ and $x=0.1$. The color coding is as in (a) and (b). Different regions in the spectra are marked with the bold case letters. \textbf{a} corresponds to the elastic line, \textbf{b} to the energy region from approximately 30 meV to 100 meV and \textbf{c} to the clearly resolved peak near 150 meV.  
  (d) The extracted temperature dependence of peak \textbf{c} for $x=0$ and $x=0.1$. The fitting model is explained in the body text.}
\end{figure*}

In this section we focus on the intra-t$_{2g}$ excitations, which are found below 200 meV energy loss. They are best 
resolved when exciting the system at feature A of  Fig. \ref{fig:holedoping} (a) [see the inset of Fig. \ref{fig:holedoping}(d)]. 

To get more insight concerning these low energy excitations and to relate them to the phase diagram of the 
AVO$_3$ vanadates, we acquired high-resolution data as a function of temperature from 30 K to room temperature. 
We have used the state-of-the-art energy resolution of 40 meV. Data for LVO and LSVO are presented in Fig. \ref{fig:result_tdep}(a)-(c). In Fig. \ref{fig:result_tdep} (c) present a comparison between LVO and LSVO at the 
endpoints of the temperature series (30 K and 300 K). Near the elastic line (labeled as \textbf{a}) we reveal the 
presence of an asymmetric peak (\textbf{c}) which has a maximum in the 140-160 meV range and a long tail that 
reaches the baseline near 500 meV. Furthermore, there are temperature dependent intensity variations in the 
30-80 meV energy range. We label this energy region as \textbf{b}. The complex temperature evolution we found 
reveals that different contributions are present with possibly differing temperature dependencies. 

In order to track the temperature evolution of feature \textbf{b} we have used a fitting model that we 
briefely describe here. Inelastic neutron scattering and Raman scattering studies show that magnon and 
phonon modes are expected in the \textbf{b} region.\cite{MIYASAKA2006,TUNG2008,VREJOIU2016} 
Hence we use resolution limited Pearson VII functions to capture \textbf{a} and \textbf{b} and a 
skewed Gaussian function to represent \textbf{c}. We tested for the necessity of introducing a 
peak for \textbf{b} by trying to reproduce the temperature evolution of peak \textbf{c} by a 
simple blueshifting, using its lineshape at 30 K, but we could not obtain a satisfactory fit 
indicating that our data are not consistent with a single blueshifting asymmetric peak. Therefore, 
a peak at \textbf{b} is necessary for the fit and we assume that there are unresolved and/or 
intrinsically broad excitations contributing to the spectrum and to the lineshape of \textbf{c} with temperature. 
We allowed the \textbf{b} peak energy loss to vary in the range of 20-50 meV in the final results. 
All parameters of the skewed Gaussian component were allowed to vary.  

The temperature dependence of the maximum of \textbf{c} is presented in Fig. \ref{fig:result_tdep}(d). 
We can identify two distinct regions in the temperature behaviour. From 300 K to 140 K the position of the 
maximum of \textbf{c} can be considered temperature independent within the error bars for both compounds. 
At lower temperatures, the maximum starts to blueshift. For LSVO the energy loss reaches a maximum at 100 K 
while for LVO \textbf{c} continues to blueshift down to the minimum temperature of 30 K. It is interesting 
to remark that in La$_{1-x}$Sr$_x$VO$_3$ single crystals 140 K and 100 K are the temperatures where both 
magnetic and structural transitions occur in LVO and LSVO respectively. Ellipsometry measurements show that 
the OO is maintained in strained LVO/STO films and it occours at T$_{OO}$ = 135 K. That said, pinpointing 
the origin of the difference in the T evolution between LVO and LSVO would benefit from a more thorough 
understanding of the magnetic and structural properties of strained films as a function of Sr to La substitution. 
We will discuss the origin of this temperature behaviour in section \ref{ssec:disc} in the light of the crystal 
field calculations presented in section \ref{ssec:cft}.

A high resolution RIXS study on YVO also observed low energy excitations at 100-200 meV.\cite{BENCKISER2013} 
Consistently to our data, an overall shift of 30-40 meV between room temperature and 30 K was observed.\cite{BENCKISER2013} 
For NdVO$_3$\cite{LAVEROCK2014} a temperature dependent peak was observed at 400 meV energy loss when exciting the 
RIXS near the maximum of the V L$_3$ absorption into dominantly $e_g$ symmetry states and it was emphasized how 
this feature was not found in YVO. The feature disappeared upon cooling from room temperature to 100 K and it 
was assigned to a bi-orbiton based on its high excitation energy.\cite{LAVEROCK2014} 
LaTiO$_3$ and YTiO$_3$ have also been studied with RIXS at the Ti L$_3$ edge and broad peaks attributed to 
orbital excitations were observed at an energy loss of slightly less than 300 meV.\cite{ULRICH2008,ULRICH2009} 
Large intensity variations were observed between room temperature and 20 K in an experiment performed with 
comparable resolution but in linear horizontal polarization.\cite{ULRICH2009} In this case, the excitations 
did not exhibit such clear energy shifts as observed here, but only a 10 meV shift can be deduced by visual 
inspection, hence our data pinpoint an interesting distinction between the vanadate and the titanate compounds.

\subsection{Crystal field theory calculations} \label{ssec:cft}

Crystal field multiplet calculation were performed using the Quanty\cite{HAVERKORT2012} package in order to 
interpret our results in a local picture. The simulations were done assuming a V 3$d^2$ ion. The model Hamiltonian 
included the crystal field term, valence and core-valence spin-orbit interaction and the valence and core-valence 
Coulomb interaction. The matrix elements of the Coulomb interaction, and the spin orbit coupling constants were taken 
from atomic Hartree-Fock calculations.\cite{HAVERKORT2004} 

\begin{table}[h!]
\caption{\label{tab:cf}The crystal field multiplets of a 3d$^2$ configuration in a D$_{4h}$ crystal field. The listed energies correspond to the lowest spin-orbit sublevel of the configuration further 
characterized by the irreducible representation. We also show the corresponding electron configuration in the O$_h$ nomenclature.}  
\begin{ruledtabular}
\begin{tabular}{c c c c c}
  Energy (eV) & $N_{states}$ & Irrep. & Spin & Config.\\
  \hline
    0 & 6 & $^3E$ & 1 & $^2t_{2g}$\\
 \hline
  0.163 & 3 & $^3A_2$ & 1 & $^2t_{2g}$\\
  \hline
  1.205 & 1 & $^1A_1$ & 0 & $^2t_{2g}$\\
   \hline
  1.245 & 2 & $^1E$ & 0 & $^2t_{2g}$\\
   \hline
  1.349 & 2 & $^1B_1,^1B_2 $ & 0 & $^2t_{2g}$\\
   \hline
  1.796 & 3 & $^3A_2$ & 1 & $^1t_{2g}^1e_g$\\
   \hline
   1.861 & 6 & $^3E$ & 1 & $^1t_{2g}^1e_g$\\
   \hline
    2.725 & 1 & $^1A_1$ & 0 & $^2t_{2g}$\\
   \hline
    2.830 & 3 & $^3B_2$ & 1 & $^1t_{2g}{^1}e_g$\\
   \hline
    3.061 & 6 & $^3E$ & 1 & $^1t_{2g}^1e_g$\\
   \hline
    3.207 & 1 & $^1A_2$ & 0 & $^1t_{2g}^1e_g$\\
   \hline
    3.273 & 3 & $^1E$ & 0 & $^1t_{2g}^1e_g$\\
   
\end{tabular}
\end{ruledtabular}
\end{table}

The adjustable parameters of the model in D$_{4h}$ symmetry are the screening of the Coulomb interaction $\beta$, 
the octahedral component of the crystal field splitting $Dq$ and  $Ds$ and $Dt$ that describe the distortion of 
the octahedron and the resulting intra-t$_{2g}$/e$_g$ splitting.\cite{DEGROOT2008} We also chose to use the 
same $\beta$ for the core-valence and valence Coulomb interaction. The temperature dependence of the spectra 
were simulated by using the Boltzmann distribution to calculate the statistical weights of the low energy 
eigenstates of the Hamiltonian. The weights were then used to calculate the spectrum as an average over the 
thermally populated initial electronic configurations. 

\begin{figure*}
\includegraphics[width=0.95\textwidth]{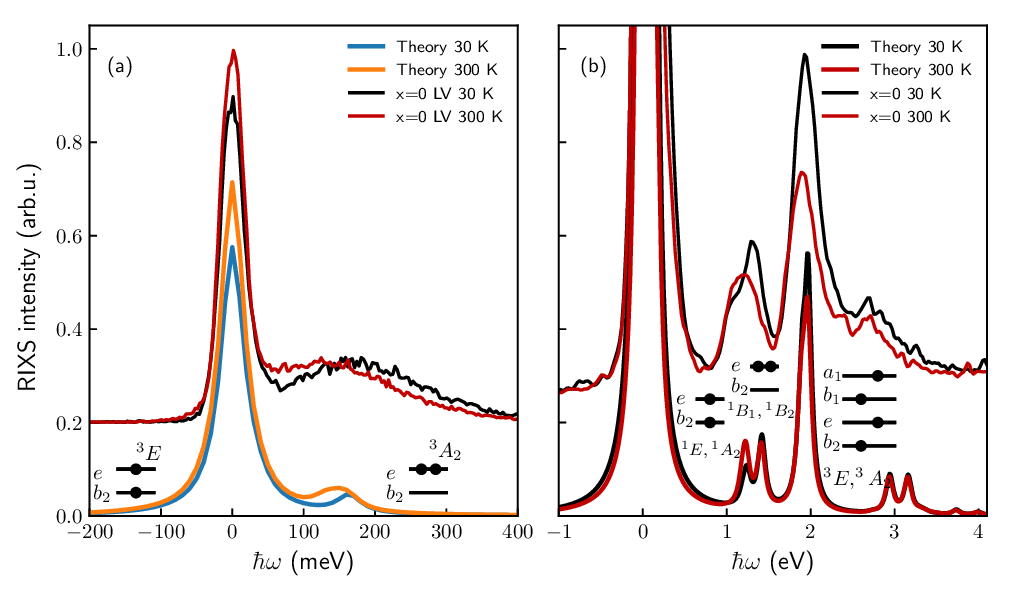}
\caption{\label{fig:cfrixs} a) High resolution experiment for LVO at 30 K vs a simulated RIXS spectrum. 
The incident energy was set feature A of Fig. \ref{fig:holedoping} (a). Due to the domain structure of 
the sample, the presented spectra are averages over the appropriate linear polarization configurations. b) 
RIXS simulated at feature B of Fig. \ref{fig:holedoping}. The energy level scheme for the $^3A_2$/$^3E$ peak 
is to be understood as possible configurations of $^1t_{2g}{^1e_g}$ with S=1.}
\end{figure*}

By comparison with the experimental XAS in Fig.\ref{fig:holedoping} (a), it was found 
that $\beta$ = 0.65 and $Dq=$0.21 eV do a good job in reproducing the overall lineshape, 
and also the energy separation of the peaks A-C. We found that the absorption spectrum was 
not very sensitive to the crystal field parameters $Dt$ and $Ds$. To determine them, we made 
the assumption that the RIXS peak \textbf{c} reflects the splitting of the t$_{2g}$ single particle 
orbitals into the e$_g$ and b$_{2g}$ orbitals. The splitting of the single particle states 
$\Delta E_{t_{2g}} = E_{e_g}-E_{b_{2g}} = 3 Ds - 5 Dt$ was fixed to 160 meV. We then varied 
the e$_g$ orbital splitting $\Delta E_{e{_g}} = E_{b_{1g}} - E_{a_{1g}} = 4 Ds + 5 Dt$ under 
the constraint $3 Ds - 5 Dt = 160 $ meV, computed RIXS spectra and compared the results to experiment in Fig. \ref{fig:holedoping}(c). We found that $Ds$ = -0.037 eV and $Dt$ = 0.009 eV give a good 
account of the RIXS spectra. During the comparisons, it was also found that only small 
values $|Ds| < 50$ meV and $|Dt| < 50$ meV of the distortion parameters resulted in a 
lineshape in qualitative agreement with experiment. We show these results for incident energies 
corresponding to features A and B of  
Fig. \ref{fig:holedoping}(a) in \ref{fig:cfrixs}(a) and (b), respectively. The good agreement verifies that the distortion of the 
octahedron in LVO is indeed quite small, and it is in line with results of LDA+DMFT calculations on the 
crystal field levels of LVO and with structural data.\cite{BORDET1993,SOLOVYEV2006} 
The LDA+DMFT calculations had been performed for the bulk monoclinic structure, and the 
V(1) site was found to have intra-t$_{2g}$ crystal field splittings  of 50 meV and approximately 
150 meV, and the V(2) site had 50 meV and approximatively 200 meV. 

We present the obtained states of the 3d$^2$ configuration up to 3.4 eV from the ground state 
term in Table \ref{tab:cf}. The first states are $S$=1 and they are associated with the $O_h$  $^3$T$_1$ 
ground state term that are split to $^3E$ and $^3A_2$ in D$_{4h}$ with the former being the ground state. \cite{YUSHANKAI2013} 
The $^3{A}_2$ term is the source of the low energy peak in the crystal field model. The next terms are $S$=0 
states ${}^1$T$_2$, ${}^1$E, ${}^1$B$_1$ and ${}^1$B$_2$, of which the two former correspond to the low energy 
side and the two latter to the high energy side of the double peak observed in the experiment between 1.1 eV 
and 1.3 eV. This peak was assigned using the O$_h$ irreducible representations in Ref.\cite{BENCKISER2013}, 
but this simplification hides the fact that one may in fact estimate the magnitude of the intra-t$_{2g}$ 
splitting by extracting the energy difference of two peaks. The $O_h$ $^3$T$_2$ term is also split into 
components reflecting the intra-t$_{2g}$ and e${_g}$ splitting. It is found at approximately 2 eV above 
the ground state, and it is the first excited state with an $^1t_{2g}$ $^1e_g$ configuration. The assignments 
are in agreement with earlier works on NdVO$_3$ and YVO.\cite{YUSHANKAI2013,LAVEROCK2014,BENCKISER2013}  
We have also performed calculations taking metal-ligand hybridization\cite{HAVERKORT2012} into account. 
We used a charge transfer energy of 4 eV and t$_{2g}$ (e$_g$) hopping of 2.08 eV (3.12 eV) derived from literature.\cite{MOSSANEK2009} No essential differences were found in terms of the lineshape of the low 
and high energy crystal field excitations, but the Dq value had to be reduced to 0.16-0.18 eV to achieve 
good agreement with the energy position of the $^3A_2$ and $^3E$ states. The crystal field model has less 
adjustable parameters and hence we show only these results. 

\subsection{On the RIXS lineshape at the V L$_3$ edge and the origin of the temperature effect}\label{ssec:disc}

The crystal field model presented in the previous section performed well in reproducing the energies of the 
high energy crystal field excitations and was able to describe qualitatively the temperature dependence 
of their intensities. Furthermore, the model can also qualitatively describe the gap closing effect 
seen at \textbf{b}. It however fails in reproducing the evolution of the spectra at the high energy 
edge of \textbf{c}, and the intensity past 200 meV is naturally not reproduced as the intra t$_{2g}$-splitting 
sets the highest excitation energy scale in the vicinity of the elastic line. Furthermore, had we 
used e.g. $D_{3h}$ symmetry to fully split the t$_{2g}$ states along the lines of the previously cited 
LDA+DMFT results, the added excitations would contribute below 100 meV energy  losses.\cite{SOLOVYEV2006} 
The high energy continuum was attributed to biorbitons and multiphonon excitations for YVO.\cite{BENCKISER2013}   

Structural data for bulk LVO shows that the V-O bonds contract by less than 0.4\% upon cooling from room temperature to 150 K. 
Upon cooling further to 10 K, two of the bonds contract by 2.5\% and one by 1\%. \cite{BORDET1993}  
The O ions are the dominant contribution to the crystal field potential and the accelerated evolution of the V-O distances 
below the OO/AFM transition is consistent with our results.  Structural properties of LVO films on STO substrates 
have been studied with diffraction techniques.\cite{LEKSHMI2005,ROTELLA2012,ROTELLA2015,MELEY2018} Temperature 
dependence of the out of plane lattice parameter ([110] of the bulk $Pnma$ structure) has been studied using 
laboratory based X-ray diffraction and it was found to remain constant between 300 K and 10 K.\cite{MELEY2018} 
The in-plane lattice parameters are expected to follow the thermal contraction of the substrate, which 
reduces from 3.92 {\AA} to 3.91 {\AA} between 300 K and 150 K, and remains constant from thereon.\cite{MELEY2018} 
This in contrast with the behavior of bulk LVO, where the lattice parameters evolve more rapidly below the OO/AFM transition.    

The epitaxial constraint must certainly force the lattice to relax in a distinct fashion with respect to the bulk 
system when entering the ordered phase. Unfortunately refined structures of LVO on STO near and below the OO/AFM 
transition have not been published to our knowledge. It is plausible that the structural distortions associated 
with the OO and AFM transitions require high-resolution diffraction or extended X-Ray absorption fine structure 
studies to be observed. Comparing our case to bulk YVO, which evolves structurally similarly to bulk LVO between 
300 K and 100 K, a blueshift of a similar magnitude was observed between 300 K and 100K.\cite{BLAKE2002,BENCKISER2013} 
This constrains the role of the lattice distortion in the blueshift to at least mimick the bulk behavior. 
We note that a recent resonant photoemission study of LVO on LaAlO$_3$ (corresponding to compressive strain 
in contrast with the tensile strain in the case of STO) found temperature dependent features in the O dominated 
part of the valence band and interpreted their results via differences in hybridization induced by the structural 
phase transition found in bulk LVO.\cite{jana2020} 
Direct observations of the structural phase transition in LVO/LSVO films are lacking, but as the hybridization 
determines the splitting of the t$_{2g}$ states\cite{SOLOVYEV2006,PAVARINI2005} our observation of the blueshift 
is in agreement with Jana et al.\cite{jana2020} 

Raman scattering studies on LVO and PrVO$_3$ thin films have found that phonon modes associated with the bulk orbital 
ordering transition become active around 140 K in thin films as well.\cite{VREJOIU2016,VREJOIU2017} For PrVO$_3$ thin 
films, the antiferromagnetic transition has been observed and the critical temperature was observed to be reduced by 
80 K from the bulk value, whereas for LVO the T$_c$ does not appear to change.\cite{COPIE2013,WANG2015} Here, the curves 
presented in Fig. \ref{fig:result_tdep} (d) start their upturns near the bulk OO/N\'eel transition critical temperatures, 
but change is rather gradual. This could be rationalized in terms of the OO and N\'eel transition temperatures being 
separated by the substrate imposed strain creating a larger temperature range in which structural evolution takes place. 
Furthermore, we note that replacing La with a smaller R ion, separates the transitions in the bulk AVO$_3$ compounds as well. \cite{MIYASAKA2003} 
Coupling of the crystal field excitations to magnons and phonons can also have an effect. 
Magnon coupling has been studied from the orbital superexchange perspective include, where 
is was found that an orbiton excitation would redshift upon entering the AFM phase for $\textbf{q}$ 
along the ${\Gamma}X$ direction and it would redshift near $\Gamma$ and Z, and remain unchanged at the 
midpoint.\cite{ISHIHARA2004} Assuming an orbital superexchange parameter of 40 meV following Ref. \cite{BENCKISER2013}, 
the model predicts a redshift of 10-20 meV. Our data shows no evidence for a red shifting low energy excitation when 
entering the AFM phase. LaTiO$_3$ and YTiO$_3$ have similar magnetic properties as the corresponding vanadates, but 
in their case the shift between room temperature and the AFM/OO phase is of the order 10 meV. However, 
for LaTiO$_3$ and YTiO$_3$ the Ti-O bonds change by a few parts per thousand between room temperature and 10 K, whereas LVO and YVO exhibit changes of the order 1\%. 

Hence we conclude that the evolution of the crystal field splitting driven by the AFM/OO transition is the dominant contribution 
to the blueshifting of the low energy excitation spectrum. It has been suggested that the orbital superexchange mechanism 
is active at room temperature in bulk LVO, and is lost rapidly upon cooling\cite{DeRaychaudhury2007}. We find that 
the spectral features resolved in the present RIXS experiment do not change appreciably until reaching the AFM/OO phase 
boundary. The momentum transfer dependence of the RIXS signal should be studied for a conclusive proof.  

\section{Conclusions}

We have presented a RIXS study of temperature and doping evolution of low energy excitations in LVO and LSVO. 
We found that intra-t$_{2g}$ excitations below 200 meV blueshift upon entering the AFM/OO phase. We argue that 
the shift reflects mainly the evolution of the nearest- and next-to-nearest atomic positions about the V ions. 
The stability of the peak position between room temperature and the AFM/OO phase boundary signifies that the 
proposed orbital fluctuations are not important for determining the energy of this excitation. Our crystal 
field calculations provide a good account of the XAS and RIXS spectra and support a local interpretation. 
Furthermore, we find no clear signatures of V ions changing their valency upon light hole doping under bulk 
sensitive experimental conditions which implies that the MIT in LVO and LSVO deviates from a simple Mott-Hubbard 
or charge-transfer pictures. Our finding thus calls for a detailed theoretical investigation of doping effects 
beyond model approaches. Finally, we have shown that phonon response is responsible of the low energy excitation 
spectrum observed in the O pre-edge region.

\begin{acknowledgements} 
This work has been supported by the Labex Palm (Grant No.ANR-10-LABX-0039-PALM).
Carnot ESP Funded by Agence Nationale de la Recherche.
We acknowledge Synchrotron SOLEIL and the ESRF for providing synchrotron radiation and technical support. 
\end{acknowledgements}

\bibliography{apstemplate_v1.bib}

\begin{thebibliography}{95}%
\makeatletter
\providecommand \@ifxundefined [1]{%
 \@ifx{#1\undefined}
}%
\providecommand \@ifnum [1]{%
 \ifnum #1\expandafter \@firstoftwo
 \else \expandafter \@secondoftwo
 \fi
}%
\providecommand \@ifx [1]{%
 \ifx #1\expandafter \@firstoftwo
 \else \expandafter \@secondoftwo
 \fi
}%
\providecommand \natexlab [1]{#1}%
\providecommand \enquote  [1]{``#1''}%
\providecommand \bibnamefont  [1]{#1}%
\providecommand \bibfnamefont [1]{#1}%
\providecommand \citenamefont [1]{#1}%
\providecommand \href@noop [0]{\@secondoftwo}%
\providecommand \href [0]{\begingroup \@sanitize@url \@href}%
\providecommand \@href[1]{\@@startlink{#1}\@@href}%
\providecommand \@@href[1]{\endgroup#1\@@endlink}%
\providecommand \@sanitize@url [0]{\catcode `\\12\catcode `\$12\catcode
  `\&12\catcode `\#12\catcode `\^12\catcode `\_12\catcode `\%12\relax}%
\providecommand \@@startlink[1]{}%
\providecommand \@@endlink[0]{}%
\providecommand \url  [0]{\begingroup\@sanitize@url \@url }%
\providecommand \@url [1]{\endgroup\@href {#1}{\urlprefix }}%
\providecommand \urlprefix  [0]{URL }%
\providecommand \Eprint [0]{\href }%
\providecommand \doibase [0]{http://dx.doi.org/}%
\providecommand \selectlanguage [0]{\@gobble}%
\providecommand \bibinfo  [0]{\@secondoftwo}%
\providecommand \bibfield  [0]{\@secondoftwo}%
\providecommand \translation [1]{[#1]}%
\providecommand \BibitemOpen [0]{}%
\providecommand \bibitemStop [0]{}%
\providecommand \bibitemNoStop [0]{.\EOS\space}%
\providecommand \EOS [0]{\spacefactor3000\relax}%
\providecommand \BibitemShut  [1]{\csname bibitem#1\endcsname}%
\let\auto@bib@innerbib\@empty
\bibitem [{\citenamefont {Sage}(2006)}]{Sagephd}%
  \BibitemOpen
  \bibfield  {author} {\bibinfo {author} {\bibfnamefont {M.-H.}\ \bibnamefont
  {Sage}},\ }\href
  {https://www.rug.nl/research/portal/files/2880868/10_thesis.pdf} {Ph.D.
  thesis},\ \bibinfo  {school} {University of Groningen} (\bibinfo {year}
  {2006})\BibitemShut {NoStop}%
\bibitem [{\citenamefont {Miyasaka}\ \emph {et~al.}(2003)\citenamefont
  {Miyasaka}, \citenamefont {Okimoto}, \citenamefont {Iwama},\ and\
  \citenamefont {Tokura}}]{MIYASAKA2003}%
  \BibitemOpen
  \bibfield  {author} {\bibinfo {author} {\bibfnamefont {S.}~\bibnamefont
  {Miyasaka}}, \bibinfo {author} {\bibfnamefont {Y.}~\bibnamefont {Okimoto}},
  \bibinfo {author} {\bibfnamefont {M.}~\bibnamefont {Iwama}}, \ and\ \bibinfo
  {author} {\bibfnamefont {Y.}~\bibnamefont {Tokura}},\ }\href {\doibase
  10.1103/PhysRevB.68.100406} {\bibfield  {journal} {\bibinfo  {journal} {Phys.
  Rev. B}\ }\textbf {\bibinfo {volume} {68}},\ \bibinfo {pages} {100406(R)}
  (\bibinfo {year} {2003})}\BibitemShut {NoStop}%
\bibitem [{\citenamefont {Mizokawa}\ and\ \citenamefont
  {Fujimori}(1996)}]{Mizokawa1996}%
  \BibitemOpen
  \bibfield  {author} {\bibinfo {author} {\bibfnamefont {T.}~\bibnamefont
  {Mizokawa}}\ and\ \bibinfo {author} {\bibfnamefont {A.}~\bibnamefont
  {Fujimori}},\ }\href {\doibase 10.1103/PhysRevB.54.5368} {\bibfield
  {journal} {\bibinfo  {journal} {Phys. Rev. B}\ }\textbf {\bibinfo {volume}
  {54}},\ \bibinfo {pages} {5368} (\bibinfo {year} {1996})}\BibitemShut
  {NoStop}%
\bibitem [{\citenamefont {Sawada}\ \emph {et~al.}(1996)\citenamefont {Sawada},
  \citenamefont {Hamada}, \citenamefont {Terakura},\ and\ \citenamefont
  {Asada}}]{Sawada1996}%
  \BibitemOpen
  \bibfield  {author} {\bibinfo {author} {\bibfnamefont {H.}~\bibnamefont
  {Sawada}}, \bibinfo {author} {\bibfnamefont {N.}~\bibnamefont {Hamada}},
  \bibinfo {author} {\bibfnamefont {K.}~\bibnamefont {Terakura}}, \ and\
  \bibinfo {author} {\bibfnamefont {T.}~\bibnamefont {Asada}},\ }\href
  {\doibase 10.1103/PhysRevB.53.12742} {\bibfield  {journal} {\bibinfo
  {journal} {Phys. Rev. B}\ }\textbf {\bibinfo {volume} {53}},\ \bibinfo
  {pages} {12742} (\bibinfo {year} {1996})}\BibitemShut {NoStop}%
\bibitem [{\citenamefont {Mizokawa}\ \emph {et~al.}(1999)\citenamefont
  {Mizokawa}, \citenamefont {Khomskii},\ and\ \citenamefont
  {Sawatzky}}]{Mizokawa1999}%
  \BibitemOpen
  \bibfield  {author} {\bibinfo {author} {\bibfnamefont {T.}~\bibnamefont
  {Mizokawa}}, \bibinfo {author} {\bibfnamefont {D.~I.}\ \bibnamefont
  {Khomskii}}, \ and\ \bibinfo {author} {\bibfnamefont {G.~A.}\ \bibnamefont
  {Sawatzky}},\ }\href {\doibase 10.1103/PhysRevB.60.7309} {\bibfield
  {journal} {\bibinfo  {journal} {Phys. Rev. B}\ }\textbf {\bibinfo {volume}
  {60}},\ \bibinfo {pages} {7309} (\bibinfo {year} {1999})}\BibitemShut
  {NoStop}%
\bibitem [{\citenamefont {Khaliullin}\ \emph {et~al.}(2001)\citenamefont
  {Khaliullin}, \citenamefont {Horsch},\ and\ \citenamefont
  {Ole\ifmmode~\acute{s}\else \'{s}\fi{}}}]{Khaliullin2001}%
  \BibitemOpen
  \bibfield  {author} {\bibinfo {author} {\bibfnamefont {G.}~\bibnamefont
  {Khaliullin}}, \bibinfo {author} {\bibfnamefont {P.}~\bibnamefont {Horsch}},
  \ and\ \bibinfo {author} {\bibfnamefont {A.~M.}\ \bibnamefont
  {Ole\ifmmode~\acute{s}\else \'{s}\fi{}}},\ }\href {\doibase
  10.1103/PhysRevLett.86.3879} {\bibfield  {journal} {\bibinfo  {journal}
  {Phys. Rev. Lett.}\ }\textbf {\bibinfo {volume} {86}},\ \bibinfo {pages}
  {3879} (\bibinfo {year} {2001})}\BibitemShut {NoStop}%
\bibitem [{\citenamefont {Motome}\ \emph {et~al.}(2003)\citenamefont {Motome},
  \citenamefont {Seo}, \citenamefont {Fang},\ and\ \citenamefont
  {Nagaosa}}]{Motome2003}%
  \BibitemOpen
  \bibfield  {author} {\bibinfo {author} {\bibfnamefont {Y.}~\bibnamefont
  {Motome}}, \bibinfo {author} {\bibfnamefont {H.}~\bibnamefont {Seo}},
  \bibinfo {author} {\bibfnamefont {Z.}~\bibnamefont {Fang}}, \ and\ \bibinfo
  {author} {\bibfnamefont {N.}~\bibnamefont {Nagaosa}},\ }\href {\doibase
  10.1103/PhysRevLett.90.146602} {\bibfield  {journal} {\bibinfo  {journal}
  {Phys. Rev. Lett.}\ }\textbf {\bibinfo {volume} {90}},\ \bibinfo {pages}
  {146602} (\bibinfo {year} {2003})}\BibitemShut {NoStop}%
\bibitem [{\citenamefont {Ren}\ \emph {et~al.}(2003)\citenamefont {Ren},
  \citenamefont {Nugroho}, \citenamefont {Menovsky}, \citenamefont {Strempfer},
  \citenamefont {R\"utt}, \citenamefont {Iga}, \citenamefont {Takabatake},\
  and\ \citenamefont {Kimball}}]{REN2003}%
  \BibitemOpen
  \bibfield  {author} {\bibinfo {author} {\bibfnamefont {Y.}~\bibnamefont
  {Ren}}, \bibinfo {author} {\bibfnamefont {A.~A.}\ \bibnamefont {Nugroho}},
  \bibinfo {author} {\bibfnamefont {A.~A.}\ \bibnamefont {Menovsky}}, \bibinfo
  {author} {\bibfnamefont {J.}~\bibnamefont {Strempfer}}, \bibinfo {author}
  {\bibfnamefont {U.}~\bibnamefont {R\"utt}}, \bibinfo {author} {\bibfnamefont
  {F.}~\bibnamefont {Iga}}, \bibinfo {author} {\bibfnamefont {T.}~\bibnamefont
  {Takabatake}}, \ and\ \bibinfo {author} {\bibfnamefont {C.~W.}\ \bibnamefont
  {Kimball}},\ }\href {\doibase 10.1103/PhysRevB.67.014107} {\bibfield
  {journal} {\bibinfo  {journal} {Phys. Rev. B}\ }\textbf {\bibinfo {volume}
  {67}},\ \bibinfo {pages} {014107} (\bibinfo {year} {2003})}\BibitemShut
  {NoStop}%
\bibitem [{\citenamefont {Ulrich}\ \emph {et~al.}(2003)\citenamefont {Ulrich},
  \citenamefont {Khaliullin}, \citenamefont {Sirker}, \citenamefont {Reehuis},
  \citenamefont {Ohl}, \citenamefont {Miyasaka}, \citenamefont {Tokura},\ and\
  \citenamefont {Keimer}}]{Ulrich2003}%
  \BibitemOpen
  \bibfield  {author} {\bibinfo {author} {\bibfnamefont {C.}~\bibnamefont
  {Ulrich}}, \bibinfo {author} {\bibfnamefont {G.}~\bibnamefont {Khaliullin}},
  \bibinfo {author} {\bibfnamefont {J.}~\bibnamefont {Sirker}}, \bibinfo
  {author} {\bibfnamefont {M.}~\bibnamefont {Reehuis}}, \bibinfo {author}
  {\bibfnamefont {M.}~\bibnamefont {Ohl}}, \bibinfo {author} {\bibfnamefont
  {S.}~\bibnamefont {Miyasaka}}, \bibinfo {author} {\bibfnamefont
  {Y.}~\bibnamefont {Tokura}}, \ and\ \bibinfo {author} {\bibfnamefont
  {B.}~\bibnamefont {Keimer}},\ }\href {\doibase 10.1103/PhysRevLett.91.257202}
  {\bibfield  {journal} {\bibinfo  {journal} {Phys. Rev. Lett.}\ }\textbf
  {\bibinfo {volume} {91}},\ \bibinfo {pages} {257202} (\bibinfo {year}
  {2003})}\BibitemShut {NoStop}%
\bibitem [{\citenamefont {Fang}\ and\ \citenamefont
  {Nagaosa}(2004)}]{Fang2004}%
  \BibitemOpen
  \bibfield  {author} {\bibinfo {author} {\bibfnamefont {Z.}~\bibnamefont
  {Fang}}\ and\ \bibinfo {author} {\bibfnamefont {N.}~\bibnamefont {Nagaosa}},\
  }\href {\doibase 10.1103/PhysRevLett.93.176404} {\bibfield  {journal}
  {\bibinfo  {journal} {Phys. Rev. Lett.}\ }\textbf {\bibinfo {volume} {93}},\
  \bibinfo {pages} {176404} (\bibinfo {year} {2004})}\BibitemShut {NoStop}%
\bibitem [{\citenamefont {Miyasaka}\ \emph {et~al.}(2005)\citenamefont
  {Miyasaka}, \citenamefont {Onoda}, \citenamefont {Okimoto}, \citenamefont
  {Fujioka}, \citenamefont {Iwama}, \citenamefont {Nagaosa},\ and\
  \citenamefont {Tokura}}]{Miyasaka2005}%
  \BibitemOpen
  \bibfield  {author} {\bibinfo {author} {\bibfnamefont {S.}~\bibnamefont
  {Miyasaka}}, \bibinfo {author} {\bibfnamefont {S.}~\bibnamefont {Onoda}},
  \bibinfo {author} {\bibfnamefont {Y.}~\bibnamefont {Okimoto}}, \bibinfo
  {author} {\bibfnamefont {J.}~\bibnamefont {Fujioka}}, \bibinfo {author}
  {\bibfnamefont {M.}~\bibnamefont {Iwama}}, \bibinfo {author} {\bibfnamefont
  {N.}~\bibnamefont {Nagaosa}}, \ and\ \bibinfo {author} {\bibfnamefont
  {Y.}~\bibnamefont {Tokura}},\ }\href {\doibase 10.1103/PhysRevLett.94.076405}
  {\bibfield  {journal} {\bibinfo  {journal} {Phys. Rev. Lett.}\ }\textbf
  {\bibinfo {volume} {94}},\ \bibinfo {pages} {076405} (\bibinfo {year}
  {2005})}\BibitemShut {NoStop}%
\bibitem [{\citenamefont {Solovyev}(2006)}]{SOLOVYEV2006}%
  \BibitemOpen
  \bibfield  {author} {\bibinfo {author} {\bibfnamefont {I.~V.}\ \bibnamefont
  {Solovyev}},\ }\href {\doibase 10.1103/PhysRevB.74.054412} {\bibfield
  {journal} {\bibinfo  {journal} {Phys. Rev. B}\ }\textbf {\bibinfo {volume}
  {74}},\ \bibinfo {pages} {054412} (\bibinfo {year} {2006})}\BibitemShut
  {NoStop}%
\bibitem [{\citenamefont {De~Raychaudhury}\ \emph {et~al.}(2007)\citenamefont
  {De~Raychaudhury}, \citenamefont {Pavarini},\ and\ \citenamefont
  {Andersen}}]{DeRaychaudhury2007}%
  \BibitemOpen
  \bibfield  {author} {\bibinfo {author} {\bibfnamefont {M.}~\bibnamefont
  {De~Raychaudhury}}, \bibinfo {author} {\bibfnamefont {E.}~\bibnamefont
  {Pavarini}}, \ and\ \bibinfo {author} {\bibfnamefont {O.~K.}\ \bibnamefont
  {Andersen}},\ }\href {\doibase 10.1103/PhysRevLett.99.126402} {\bibfield
  {journal} {\bibinfo  {journal} {Phys. Rev. Lett.}\ }\textbf {\bibinfo
  {volume} {99}},\ \bibinfo {pages} {126402} (\bibinfo {year}
  {2007})}\BibitemShut {NoStop}%
\bibitem [{\citenamefont {Horsch}\ \emph {et~al.}(2008)\citenamefont {Horsch},
  \citenamefont {Ole\ifmmode~\acute{s}\else \'{s}\fi{}}, \citenamefont
  {Feiner},\ and\ \citenamefont {Khaliullin}}]{Horsch2008}%
  \BibitemOpen
  \bibfield  {author} {\bibinfo {author} {\bibfnamefont {P.}~\bibnamefont
  {Horsch}}, \bibinfo {author} {\bibfnamefont {A.~M.}\ \bibnamefont
  {Ole\ifmmode~\acute{s}\else \'{s}\fi{}}}, \bibinfo {author} {\bibfnamefont
  {L.~F.}\ \bibnamefont {Feiner}}, \ and\ \bibinfo {author} {\bibfnamefont
  {G.}~\bibnamefont {Khaliullin}},\ }\href {\doibase
  10.1103/PhysRevLett.100.167205} {\bibfield  {journal} {\bibinfo  {journal}
  {Phys. Rev. Lett.}\ }\textbf {\bibinfo {volume} {100}},\ \bibinfo {pages}
  {167205} (\bibinfo {year} {2008})}\BibitemShut {NoStop}%
\bibitem [{\citenamefont {Zhou}\ \emph {et~al.}(2008)\citenamefont {Zhou},
  \citenamefont {Ren}, \citenamefont {Yan}, \citenamefont {Mitchell},\ and\
  \citenamefont {Goodenough}}]{ZHOU2008}%
  \BibitemOpen
  \bibfield  {author} {\bibinfo {author} {\bibfnamefont {J.-S.}\ \bibnamefont
  {Zhou}}, \bibinfo {author} {\bibfnamefont {Y.}~\bibnamefont {Ren}}, \bibinfo
  {author} {\bibfnamefont {J.-Q.}\ \bibnamefont {Yan}}, \bibinfo {author}
  {\bibfnamefont {J.~F.}\ \bibnamefont {Mitchell}}, \ and\ \bibinfo {author}
  {\bibfnamefont {J.~B.}\ \bibnamefont {Goodenough}},\ }\href {\doibase
  10.1103/PhysRevLett.100.046401} {\bibfield  {journal} {\bibinfo  {journal}
  {Phys. Rev. Lett.}\ }\textbf {\bibinfo {volume} {100}},\ \bibinfo {pages}
  {046401} (\bibinfo {year} {2008})}\BibitemShut {NoStop}%
\bibitem [{\citenamefont {Ro\ifmmode~\acute{s}\else \'{s}\fi{}ciszewski}\ and\
  \citenamefont {Ole\ifmmode~\acute{s}\else
  \'{s}\fi{}}(2018)}]{Rosciszewski2018}%
  \BibitemOpen
  \bibfield  {author} {\bibinfo {author} {\bibfnamefont {K.}~\bibnamefont
  {Ro\ifmmode~\acute{s}\else \'{s}\fi{}ciszewski}}\ and\ \bibinfo {author}
  {\bibfnamefont {A.~M.}\ \bibnamefont {Ole\ifmmode~\acute{s}\else
  \'{s}\fi{}}},\ }\href {\doibase 10.1103/PhysRevB.98.085119} {\bibfield
  {journal} {\bibinfo  {journal} {Phys. Rev. B}\ }\textbf {\bibinfo {volume}
  {98}},\ \bibinfo {pages} {085119} (\bibinfo {year} {2018})}\BibitemShut
  {NoStop}%
\bibitem [{\citenamefont {Blake}\ \emph {et~al.}(2002)\citenamefont {Blake},
  \citenamefont {Palstra}, \citenamefont {Ren}, \citenamefont {Nugroho},\ and\
  \citenamefont {Menovsky}}]{BLAKE2002}%
  \BibitemOpen
  \bibfield  {author} {\bibinfo {author} {\bibfnamefont {G.~R.}\ \bibnamefont
  {Blake}}, \bibinfo {author} {\bibfnamefont {T.~T.~M.}\ \bibnamefont
  {Palstra}}, \bibinfo {author} {\bibfnamefont {Y.}~\bibnamefont {Ren}},
  \bibinfo {author} {\bibfnamefont {A.~A.}\ \bibnamefont {Nugroho}}, \ and\
  \bibinfo {author} {\bibfnamefont {A.~A.}\ \bibnamefont {Menovsky}},\ }\href
  {\doibase 10.1103/PhysRevB.65.174112} {\bibfield  {journal} {\bibinfo
  {journal} {Phys. Rev. B}\ }\textbf {\bibinfo {volume} {65}},\ \bibinfo
  {pages} {174112} (\bibinfo {year} {2002})}\BibitemShut {NoStop}%
\bibitem [{\citenamefont {Benckiser}\ \emph {et~al.}(2013)\citenamefont
  {Benckiser}, \citenamefont {Fels}, \citenamefont {Ghiringhelli},
  \citenamefont {Moretti~Sala}, \citenamefont {Schmitt}, \citenamefont
  {Schlappa}, \citenamefont {Strocov}, \citenamefont {Mufti}, \citenamefont
  {Blake}, \citenamefont {Nugroho}, \citenamefont {Palstra}, \citenamefont
  {Haverkort}, \citenamefont {Wohlfeld},\ and\ \citenamefont
  {Gr\"uninger}}]{BENCKISER2013}%
  \BibitemOpen
  \bibfield  {author} {\bibinfo {author} {\bibfnamefont {E.}~\bibnamefont
  {Benckiser}}, \bibinfo {author} {\bibfnamefont {L.}~\bibnamefont {Fels}},
  \bibinfo {author} {\bibfnamefont {G.}~\bibnamefont {Ghiringhelli}}, \bibinfo
  {author} {\bibfnamefont {M.}~\bibnamefont {Moretti~Sala}}, \bibinfo {author}
  {\bibfnamefont {T.}~\bibnamefont {Schmitt}}, \bibinfo {author} {\bibfnamefont
  {J.}~\bibnamefont {Schlappa}}, \bibinfo {author} {\bibfnamefont {V.~N.}\
  \bibnamefont {Strocov}}, \bibinfo {author} {\bibfnamefont {N.}~\bibnamefont
  {Mufti}}, \bibinfo {author} {\bibfnamefont {G.~R.}\ \bibnamefont {Blake}},
  \bibinfo {author} {\bibfnamefont {A.~A.}\ \bibnamefont {Nugroho}}, \bibinfo
  {author} {\bibfnamefont {T.~T.~M.}\ \bibnamefont {Palstra}}, \bibinfo
  {author} {\bibfnamefont {M.~W.}\ \bibnamefont {Haverkort}}, \bibinfo {author}
  {\bibfnamefont {K.}~\bibnamefont {Wohlfeld}}, \ and\ \bibinfo {author}
  {\bibfnamefont {M.}~\bibnamefont {Gr\"uninger}},\ }\href {\doibase
  10.1103/PhysRevB.88.205115} {\bibfield  {journal} {\bibinfo  {journal} {Phys.
  Rev. B}\ }\textbf {\bibinfo {volume} {88}},\ \bibinfo {pages} {205115}
  (\bibinfo {year} {2013})}\BibitemShut {NoStop}%
\bibitem [{\citenamefont {Reul}\ \emph {et~al.}(2012)\citenamefont {Reul},
  \citenamefont {Nugroho}, \citenamefont {Palstra},\ and\ \citenamefont
  {Gr\"uninger}}]{Reul2012}%
  \BibitemOpen
  \bibfield  {author} {\bibinfo {author} {\bibfnamefont {J.}~\bibnamefont
  {Reul}}, \bibinfo {author} {\bibfnamefont {A.~A.}\ \bibnamefont {Nugroho}},
  \bibinfo {author} {\bibfnamefont {T.~T.~M.}\ \bibnamefont {Palstra}}, \ and\
  \bibinfo {author} {\bibfnamefont {M.}~\bibnamefont {Gr\"uninger}},\ }\href
  {\doibase 10.1103/PhysRevB.86.125128} {\bibfield  {journal} {\bibinfo
  {journal} {Phys. Rev. B}\ }\textbf {\bibinfo {volume} {86}},\ \bibinfo
  {pages} {125128} (\bibinfo {year} {2012})}\BibitemShut {NoStop}%
\bibitem [{\citenamefont {Bordet}\ \emph {et~al.}(1993)\citenamefont {Bordet},
  \citenamefont {Chaillout}, \citenamefont {Marezio}, \citenamefont {Huang},
  \citenamefont {Santoro}, \citenamefont {Cheong}, \citenamefont {Takagi},
  \citenamefont {Oglesby},\ and\ \citenamefont {Batlogg}}]{BORDET1993}%
  \BibitemOpen
  \bibfield  {author} {\bibinfo {author} {\bibfnamefont {P.}~\bibnamefont
  {Bordet}}, \bibinfo {author} {\bibfnamefont {C.}~\bibnamefont {Chaillout}},
  \bibinfo {author} {\bibfnamefont {M.}~\bibnamefont {Marezio}}, \bibinfo
  {author} {\bibfnamefont {Q.}~\bibnamefont {Huang}}, \bibinfo {author}
  {\bibfnamefont {A.}~\bibnamefont {Santoro}}, \bibinfo {author} {\bibfnamefont
  {S.-W.}\ \bibnamefont {Cheong}}, \bibinfo {author} {\bibfnamefont
  {H.}~\bibnamefont {Takagi}}, \bibinfo {author} {\bibfnamefont
  {C.}~\bibnamefont {Oglesby}}, \ and\ \bibinfo {author} {\bibfnamefont
  {B.}~\bibnamefont {Batlogg}},\ }\href {\doibase
  https://doi.org/10.1006/jssc.1993.1285} {\bibfield  {journal} {\bibinfo
  {journal} {J. Solid State Chem.}\ }\textbf {\bibinfo {volume} {106}},\
  \bibinfo {pages} {253 } (\bibinfo {year} {1993})}\BibitemShut {NoStop}%
\bibitem [{\citenamefont {Sclauzero}\ and\ \citenamefont
  {Ederer}(2015)}]{Sclauzero2015}%
  \BibitemOpen
  \bibfield  {author} {\bibinfo {author} {\bibfnamefont {G.}~\bibnamefont
  {Sclauzero}}\ and\ \bibinfo {author} {\bibfnamefont {C.}~\bibnamefont
  {Ederer}},\ }\href {\doibase 10.1103/PhysRevB.92.235112} {\bibfield
  {journal} {\bibinfo  {journal} {Phys. Rev. B}\ }\textbf {\bibinfo {volume}
  {92}},\ \bibinfo {pages} {235112} (\bibinfo {year} {2015})}\BibitemShut
  {NoStop}%
\bibitem [{\citenamefont {Sclauzero}\ \emph {et~al.}(2016)\citenamefont
  {Sclauzero}, \citenamefont {Dymkowski},\ and\ \citenamefont
  {Ederer}}]{Sclauzero2016}%
  \BibitemOpen
  \bibfield  {author} {\bibinfo {author} {\bibfnamefont {G.}~\bibnamefont
  {Sclauzero}}, \bibinfo {author} {\bibfnamefont {K.}~\bibnamefont
  {Dymkowski}}, \ and\ \bibinfo {author} {\bibfnamefont {C.}~\bibnamefont
  {Ederer}},\ }\href {\doibase 10.1103/PhysRevB.94.245109} {\bibfield
  {journal} {\bibinfo  {journal} {Phys. Rev. B}\ }\textbf {\bibinfo {volume}
  {94}},\ \bibinfo {pages} {245109} (\bibinfo {year} {2016})}\BibitemShut
  {NoStop}%
\bibitem [{\citenamefont {He}\ \emph {et~al.}(2012)\citenamefont {He},
  \citenamefont {Sanders}, \citenamefont {Gray}, \citenamefont {Wong},
  \citenamefont {Mehta},\ and\ \citenamefont {Suzuki}}]{He2012}%
  \BibitemOpen
  \bibfield  {author} {\bibinfo {author} {\bibfnamefont {C.}~\bibnamefont
  {He}}, \bibinfo {author} {\bibfnamefont {T.~D.}\ \bibnamefont {Sanders}},
  \bibinfo {author} {\bibfnamefont {M.~T.}\ \bibnamefont {Gray}}, \bibinfo
  {author} {\bibfnamefont {F.~J.}\ \bibnamefont {Wong}}, \bibinfo {author}
  {\bibfnamefont {V.~V.}\ \bibnamefont {Mehta}}, \ and\ \bibinfo {author}
  {\bibfnamefont {Y.}~\bibnamefont {Suzuki}},\ }\href {\doibase
  10.1103/PhysRevB.86.081401} {\bibfield  {journal} {\bibinfo  {journal} {Phys.
  Rev. B}\ }\textbf {\bibinfo {volume} {86}},\ \bibinfo {pages} {081401(R)}
  (\bibinfo {year} {2012})}\BibitemShut {NoStop}%
\bibitem [{\citenamefont {Weng}\ and\ \citenamefont
  {Terakura}(2010)}]{Weng2010}%
  \BibitemOpen
  \bibfield  {author} {\bibinfo {author} {\bibfnamefont {H.}~\bibnamefont
  {Weng}}\ and\ \bibinfo {author} {\bibfnamefont {K.}~\bibnamefont
  {Terakura}},\ }\href {\doibase 10.1103/PhysRevB.82.115105} {\bibfield
  {journal} {\bibinfo  {journal} {Phys. Rev. B}\ }\textbf {\bibinfo {volume}
  {82}},\ \bibinfo {pages} {115105} (\bibinfo {year} {2010})}\BibitemShut
  {NoStop}%
\bibitem [{\citenamefont {Hotta}\ \emph {et~al.}(2007)\citenamefont {Hotta},
  \citenamefont {Susaki},\ and\ \citenamefont {Hwang}}]{Hotta2007}%
  \BibitemOpen
  \bibfield  {author} {\bibinfo {author} {\bibfnamefont {Y.}~\bibnamefont
  {Hotta}}, \bibinfo {author} {\bibfnamefont {T.}~\bibnamefont {Susaki}}, \
  and\ \bibinfo {author} {\bibfnamefont {H.~Y.}\ \bibnamefont {Hwang}},\ }\href
  {\doibase 10.1103/PhysRevLett.99.236805} {\bibfield  {journal} {\bibinfo
  {journal} {Phys. Rev. Lett.}\ }\textbf {\bibinfo {volume} {99}},\ \bibinfo
  {pages} {236805} (\bibinfo {year} {2007})}\BibitemShut {NoStop}%
\bibitem [{\citenamefont {Jana}\ \emph {et~al.}(2018)\citenamefont {Jana},
  \citenamefont {Choudhary},\ and\ \citenamefont {Phase}}]{Jana2018}%
  \BibitemOpen
  \bibfield  {author} {\bibinfo {author} {\bibfnamefont {A.}~\bibnamefont
  {Jana}}, \bibinfo {author} {\bibfnamefont {R.~J.}\ \bibnamefont {Choudhary}},
  \ and\ \bibinfo {author} {\bibfnamefont {D.~M.}\ \bibnamefont {Phase}},\
  }\href {\doibase 10.1103/PhysRevB.98.075124} {\bibfield  {journal} {\bibinfo
  {journal} {Phys. Rev. B}\ }\textbf {\bibinfo {volume} {98}},\ \bibinfo
  {pages} {075124} (\bibinfo {year} {2018})}\BibitemShut {NoStop}%
\bibitem [{\citenamefont {Rotella}\ \emph {et~al.}(2015)\citenamefont
  {Rotella}, \citenamefont {Copie}, \citenamefont {Steciuk}, \citenamefont
  {Ouerdane}, \citenamefont {Boullay}, \citenamefont {Roussel}, \citenamefont
  {Morales}, \citenamefont {David}, \citenamefont {Pautrat}, \citenamefont
  {Mercey} \emph {et~al.}}]{ROTELLA2015}%
  \BibitemOpen
  \bibfield  {author} {\bibinfo {author} {\bibfnamefont {H.}~\bibnamefont
  {Rotella}}, \bibinfo {author} {\bibfnamefont {O.}~\bibnamefont {Copie}},
  \bibinfo {author} {\bibfnamefont {G.}~\bibnamefont {Steciuk}}, \bibinfo
  {author} {\bibfnamefont {H.}~\bibnamefont {Ouerdane}}, \bibinfo {author}
  {\bibfnamefont {P.}~\bibnamefont {Boullay}}, \bibinfo {author} {\bibfnamefont
  {P.}~\bibnamefont {Roussel}}, \bibinfo {author} {\bibfnamefont
  {M.}~\bibnamefont {Morales}}, \bibinfo {author} {\bibfnamefont
  {A.}~\bibnamefont {David}}, \bibinfo {author} {\bibfnamefont
  {A.}~\bibnamefont {Pautrat}}, \bibinfo {author} {\bibfnamefont
  {B.}~\bibnamefont {Mercey}},  \emph {et~al.},\ }\href@noop {} {\bibfield
  {journal} {\bibinfo  {journal} {J. Phys.: Condens. Matter}\ }\textbf
  {\bibinfo {volume} {27}},\ \bibinfo {pages} {175001} (\bibinfo {year}
  {2015})}\BibitemShut {NoStop}%
\bibitem [{\citenamefont {Inaba}\ \emph {et~al.}(1995)\citenamefont {Inaba},
  \citenamefont {Arima}, \citenamefont {Ishikawa}, \citenamefont {Katsufuji},\
  and\ \citenamefont {Tokura}}]{Inaba1995}%
  \BibitemOpen
  \bibfield  {author} {\bibinfo {author} {\bibfnamefont {F.}~\bibnamefont
  {Inaba}}, \bibinfo {author} {\bibfnamefont {T.}~\bibnamefont {Arima}},
  \bibinfo {author} {\bibfnamefont {T.}~\bibnamefont {Ishikawa}}, \bibinfo
  {author} {\bibfnamefont {T.}~\bibnamefont {Katsufuji}}, \ and\ \bibinfo
  {author} {\bibfnamefont {Y.}~\bibnamefont {Tokura}},\ }\href {\doibase
  10.1103/PhysRevB.52.R2221} {\bibfield  {journal} {\bibinfo  {journal} {Phys.
  Rev. B}\ }\textbf {\bibinfo {volume} {52}},\ \bibinfo {pages} {R2221}
  (\bibinfo {year} {1995})}\BibitemShut {NoStop}%
\bibitem [{\citenamefont {Miyasaka}\ \emph {et~al.}(2000)\citenamefont
  {Miyasaka}, \citenamefont {Okuda},\ and\ \citenamefont
  {Tokura}}]{MIYASAKA2000}%
  \BibitemOpen
  \bibfield  {author} {\bibinfo {author} {\bibfnamefont {S.}~\bibnamefont
  {Miyasaka}}, \bibinfo {author} {\bibfnamefont {T.}~\bibnamefont {Okuda}}, \
  and\ \bibinfo {author} {\bibfnamefont {Y.}~\bibnamefont {Tokura}},\ }\href
  {\doibase 10.1103/PhysRevLett.85.5388} {\bibfield  {journal} {\bibinfo
  {journal} {Phys. Rev. Lett.}\ }\textbf {\bibinfo {volume} {85}},\ \bibinfo
  {pages} {5388} (\bibinfo {year} {2000})}\BibitemShut {NoStop}%
\bibitem [{\citenamefont {Nguyen}\ and\ \citenamefont
  {Goodenough}(1995)}]{Nguyen1995}%
  \BibitemOpen
  \bibfield  {author} {\bibinfo {author} {\bibfnamefont {H.~C.}\ \bibnamefont
  {Nguyen}}\ and\ \bibinfo {author} {\bibfnamefont {J.~B.}\ \bibnamefont
  {Goodenough}},\ }\href {\doibase 10.1103/PhysRevB.52.324} {\bibfield
  {journal} {\bibinfo  {journal} {Phys. Rev. B}\ }\textbf {\bibinfo {volume}
  {52}},\ \bibinfo {pages} {324} (\bibinfo {year} {1995})}\BibitemShut
  {NoStop}%
\bibitem [{\citenamefont {Tung}\ \emph {et~al.}(2008)\citenamefont {Tung},
  \citenamefont {Ivanov}, \citenamefont {Schefer}, \citenamefont {Lees},
  \citenamefont {Balakrishnan},\ and\ \citenamefont {Paul}}]{TUNG2008}%
  \BibitemOpen
  \bibfield  {author} {\bibinfo {author} {\bibfnamefont {L.~D.}\ \bibnamefont
  {Tung}}, \bibinfo {author} {\bibfnamefont {A.}~\bibnamefont {Ivanov}},
  \bibinfo {author} {\bibfnamefont {J.}~\bibnamefont {Schefer}}, \bibinfo
  {author} {\bibfnamefont {M.~R.}\ \bibnamefont {Lees}}, \bibinfo {author}
  {\bibfnamefont {G.}~\bibnamefont {Balakrishnan}}, \ and\ \bibinfo {author}
  {\bibfnamefont {D.~M.}\ \bibnamefont {Paul}},\ }\href {\doibase
  10.1103/PhysRevB.78.054416} {\bibfield  {journal} {\bibinfo  {journal} {Phys.
  Rev. B}\ }\textbf {\bibinfo {volume} {78}},\ \bibinfo {pages} {054416}
  (\bibinfo {year} {2008})}\BibitemShut {NoStop}%
\bibitem [{\citenamefont {Khan}\ \emph {et~al.}(2004)\citenamefont {Khan},
  \citenamefont {Bashir}, \citenamefont {Iqbal},\ and\ \citenamefont
  {Khan}}]{Khan2004}%
  \BibitemOpen
  \bibfield  {author} {\bibinfo {author} {\bibfnamefont {R.}~\bibnamefont
  {Khan}}, \bibinfo {author} {\bibfnamefont {J.}~\bibnamefont {Bashir}},
  \bibinfo {author} {\bibfnamefont {N.}~\bibnamefont {Iqbal}}, \ and\ \bibinfo
  {author} {\bibfnamefont {M.}~\bibnamefont {Khan}},\ }\href {\doibase
  https://doi.org/10.1016/j.matlet.2003.10.059} {\bibfield  {journal} {\bibinfo
   {journal} {Materials Letters}\ }\textbf {\bibinfo {volume} {58}},\ \bibinfo
  {pages} {1737 } (\bibinfo {year} {2004})}\BibitemShut {NoStop}%
\bibitem [{\citenamefont {Rotella}\ \emph {et~al.}(2012)\citenamefont
  {Rotella}, \citenamefont {L\"uders}, \citenamefont {Janolin}, \citenamefont
  {Dao}, \citenamefont {Chateigner}, \citenamefont {Feyerherm}, \citenamefont
  {Dudzik},\ and\ \citenamefont {Prellier}}]{ROTELLA2012}%
  \BibitemOpen
  \bibfield  {author} {\bibinfo {author} {\bibfnamefont {H.}~\bibnamefont
  {Rotella}}, \bibinfo {author} {\bibfnamefont {U.}~\bibnamefont {L\"uders}},
  \bibinfo {author} {\bibfnamefont {P.-E.}\ \bibnamefont {Janolin}}, \bibinfo
  {author} {\bibfnamefont {V.~H.}\ \bibnamefont {Dao}}, \bibinfo {author}
  {\bibfnamefont {D.}~\bibnamefont {Chateigner}}, \bibinfo {author}
  {\bibfnamefont {R.}~\bibnamefont {Feyerherm}}, \bibinfo {author}
  {\bibfnamefont {E.}~\bibnamefont {Dudzik}}, \ and\ \bibinfo {author}
  {\bibfnamefont {W.}~\bibnamefont {Prellier}},\ }\href {\doibase
  10.1103/PhysRevB.85.184101} {\bibfield  {journal} {\bibinfo  {journal} {Phys.
  Rev. B}\ }\textbf {\bibinfo {volume} {85}},\ \bibinfo {pages} {184101}
  (\bibinfo {year} {2012})}\BibitemShut {NoStop}%
\bibitem [{\citenamefont {Hotta}\ \emph {et~al.}(2006)\citenamefont {Hotta},
  \citenamefont {Mukunoki}, \citenamefont {Susaki}, \citenamefont {Hwang},
  \citenamefont {Fitting},\ and\ \citenamefont {Muller}}]{Hotta2006}%
  \BibitemOpen
  \bibfield  {author} {\bibinfo {author} {\bibfnamefont {Y.}~\bibnamefont
  {Hotta}}, \bibinfo {author} {\bibfnamefont {Y.}~\bibnamefont {Mukunoki}},
  \bibinfo {author} {\bibfnamefont {T.}~\bibnamefont {Susaki}}, \bibinfo
  {author} {\bibfnamefont {H.~Y.}\ \bibnamefont {Hwang}}, \bibinfo {author}
  {\bibfnamefont {L.}~\bibnamefont {Fitting}}, \ and\ \bibinfo {author}
  {\bibfnamefont {D.~A.}\ \bibnamefont {Muller}},\ }\href {\doibase
  10.1063/1.2227786} {\bibfield  {journal} {\bibinfo  {journal} {Appl. Phys.
  Lett.}\ }\textbf {\bibinfo {volume} {89}},\ \bibinfo {pages} {031918}
  (\bibinfo {year} {2006})}\BibitemShut {NoStop}%
\bibitem [{\citenamefont {Lekshmi}\ \emph {et~al.}(2005)\citenamefont
  {Lekshmi}, \citenamefont {Gayen},\ and\ \citenamefont {Hegde}}]{LEKSHMI2005}%
  \BibitemOpen
  \bibfield  {author} {\bibinfo {author} {\bibfnamefont {I.~C.}\ \bibnamefont
  {Lekshmi}}, \bibinfo {author} {\bibfnamefont {A.}~\bibnamefont {Gayen}}, \
  and\ \bibinfo {author} {\bibfnamefont {M.}~\bibnamefont {Hegde}},\ }\href
  {\doibase https://doi.org/10.1016/j.jpcs.2005.06.005} {\bibfield  {journal}
  {\bibinfo  {journal} {J. Phys. Chem. Solids}\ }\textbf {\bibinfo {volume}
  {66}},\ \bibinfo {pages} {1647 } (\bibinfo {year} {2005})}\BibitemShut
  {NoStop}%
\bibitem [{\citenamefont {Meley}\ \emph {et~al.}(2018)\citenamefont {Meley},
  \citenamefont {Karandeep}, \citenamefont {Oberson}, \citenamefont
  {de~Bruijckere}, \citenamefont {Alexander}, \citenamefont {Triscone},
  \citenamefont {Ghosez},\ and\ \citenamefont {Gariglio}}]{MELEY2018}%
  \BibitemOpen
  \bibfield  {author} {\bibinfo {author} {\bibfnamefont {H.}~\bibnamefont
  {Meley}}, \bibinfo {author} {\bibnamefont {Karandeep}}, \bibinfo {author}
  {\bibfnamefont {L.}~\bibnamefont {Oberson}}, \bibinfo {author} {\bibfnamefont
  {J.}~\bibnamefont {de~Bruijckere}}, \bibinfo {author} {\bibfnamefont
  {D.~T.~L.}\ \bibnamefont {Alexander}}, \bibinfo {author} {\bibfnamefont
  {J.-M.}\ \bibnamefont {Triscone}}, \bibinfo {author} {\bibfnamefont
  {P.}~\bibnamefont {Ghosez}}, \ and\ \bibinfo {author} {\bibfnamefont
  {S.}~\bibnamefont {Gariglio}},\ }\href {\doibase 10.1063/1.5021844}
  {\bibfield  {journal} {\bibinfo  {journal} {APL Materials}\ }\textbf
  {\bibinfo {volume} {6}},\ \bibinfo {pages} {046102} (\bibinfo {year}
  {2018})}\BibitemShut {NoStop}%
\bibitem [{\citenamefont {Wang}\ \emph {et~al.}(2015)\citenamefont {Wang},
  \citenamefont {Li}, \citenamefont {Bera}, \citenamefont {Ma}, \citenamefont
  {Jin}, \citenamefont {Yuan}, \citenamefont {Yin}, \citenamefont {David},
  \citenamefont {Chen}, \citenamefont {Wu}, \citenamefont {Prellier},
  \citenamefont {Wei},\ and\ \citenamefont {Wu}}]{WANG2015}%
  \BibitemOpen
  \bibfield  {author} {\bibinfo {author} {\bibfnamefont {L.}~\bibnamefont
  {Wang}}, \bibinfo {author} {\bibfnamefont {Y.}~\bibnamefont {Li}}, \bibinfo
  {author} {\bibfnamefont {A.}~\bibnamefont {Bera}}, \bibinfo {author}
  {\bibfnamefont {C.}~\bibnamefont {Ma}}, \bibinfo {author} {\bibfnamefont
  {F.}~\bibnamefont {Jin}}, \bibinfo {author} {\bibfnamefont {K.}~\bibnamefont
  {Yuan}}, \bibinfo {author} {\bibfnamefont {W.}~\bibnamefont {Yin}}, \bibinfo
  {author} {\bibfnamefont {A.}~\bibnamefont {David}}, \bibinfo {author}
  {\bibfnamefont {W.}~\bibnamefont {Chen}}, \bibinfo {author} {\bibfnamefont
  {W.}~\bibnamefont {Wu}}, \bibinfo {author} {\bibfnamefont {W.}~\bibnamefont
  {Prellier}}, \bibinfo {author} {\bibfnamefont {S.}~\bibnamefont {Wei}}, \
  and\ \bibinfo {author} {\bibfnamefont {T.}~\bibnamefont {Wu}},\ }\href
  {\doibase 10.1103/PhysRevApplied.3.064015} {\bibfield  {journal} {\bibinfo
  {journal} {Phys. Rev. Applied}\ }\textbf {\bibinfo {volume} {3}},\ \bibinfo
  {pages} {064015} (\bibinfo {year} {2015})}\BibitemShut {NoStop}%
\bibitem [{\citenamefont {Vrejoiu}\ \emph {et~al.}(2016)\citenamefont
  {Vrejoiu}, \citenamefont {Himcinschi}, \citenamefont {Jin}, \citenamefont
  {Jia}, \citenamefont {Raab}, \citenamefont {Engelmayer}, \citenamefont
  {Waser}, \citenamefont {Dittmann},\ and\ \citenamefont {van
  Loosdrecht}}]{VREJOIU2016}%
  \BibitemOpen
  \bibfield  {author} {\bibinfo {author} {\bibfnamefont {I.}~\bibnamefont
  {Vrejoiu}}, \bibinfo {author} {\bibfnamefont {C.}~\bibnamefont {Himcinschi}},
  \bibinfo {author} {\bibfnamefont {L.}~\bibnamefont {Jin}}, \bibinfo {author}
  {\bibfnamefont {C.-L.}\ \bibnamefont {Jia}}, \bibinfo {author} {\bibfnamefont
  {N.}~\bibnamefont {Raab}}, \bibinfo {author} {\bibfnamefont {J.}~\bibnamefont
  {Engelmayer}}, \bibinfo {author} {\bibfnamefont {R.}~\bibnamefont {Waser}},
  \bibinfo {author} {\bibfnamefont {R.}~\bibnamefont {Dittmann}}, \ and\
  \bibinfo {author} {\bibfnamefont {P.~H.~M.}\ \bibnamefont {van Loosdrecht}},\
  }\href {\doibase 10.1063/1.4945658} {\bibfield  {journal} {\bibinfo
  {journal} {APL Materials}\ }\textbf {\bibinfo {volume} {4}},\ \bibinfo
  {pages} {046103} (\bibinfo {year} {2016})}\BibitemShut {NoStop}%
\bibitem [{\citenamefont {Lindfors-Vrejoiu}\ \emph {et~al.}(2017)\citenamefont
  {Lindfors-Vrejoiu}, \citenamefont {Jin}, \citenamefont {Himcinschi},
  \citenamefont {Engelmayer}, \citenamefont {Hensling}, \citenamefont {Jia},
  \citenamefont {Waser}, \citenamefont {Dittmann},\ and\ \citenamefont
  {Loosdrecht}}]{VREJOIU2017}%
  \BibitemOpen
  \bibfield  {author} {\bibinfo {author} {\bibfnamefont {I.}~\bibnamefont
  {Lindfors-Vrejoiu}}, \bibinfo {author} {\bibfnamefont {L.}~\bibnamefont
  {Jin}}, \bibinfo {author} {\bibfnamefont {C.}~\bibnamefont {Himcinschi}},
  \bibinfo {author} {\bibfnamefont {J.}~\bibnamefont {Engelmayer}}, \bibinfo
  {author} {\bibfnamefont {F.}~\bibnamefont {Hensling}}, \bibinfo {author}
  {\bibfnamefont {C.-L.}\ \bibnamefont {Jia}}, \bibinfo {author} {\bibfnamefont
  {R.}~\bibnamefont {Waser}}, \bibinfo {author} {\bibfnamefont
  {R.}~\bibnamefont {Dittmann}}, \ and\ \bibinfo {author} {\bibfnamefont
  {P.~H.~M.}\ \bibnamefont {Loosdrecht}},\ }\href {\doibase
  10.1002/pssr.201600350} {\bibfield  {journal} {\bibinfo  {journal} {Phys.
  Status Solidi RRL}\ }\textbf {\bibinfo {volume} {11}},\ \bibinfo {pages}
  {1600350} (\bibinfo {year} {2017})}\BibitemShut {NoStop}%
\bibitem [{\citenamefont {Meley}\ \emph {et~al.}(2021)\citenamefont {Meley},
  \citenamefont {Tran}, \citenamefont {Teyssier}, \citenamefont {Krieger},
  \citenamefont {Prokscha}, \citenamefont {Suter}, \citenamefont {Salman},
  \citenamefont {Viret}, \citenamefont {van~der Marel},\ and\ \citenamefont
  {Gariglio}}]{Meley2021}%
  \BibitemOpen
  \bibfield  {author} {\bibinfo {author} {\bibfnamefont {H.}~\bibnamefont
  {Meley}}, \bibinfo {author} {\bibfnamefont {M.}~\bibnamefont {Tran}},
  \bibinfo {author} {\bibfnamefont {J.}~\bibnamefont {Teyssier}}, \bibinfo
  {author} {\bibfnamefont {J.~A.}\ \bibnamefont {Krieger}}, \bibinfo {author}
  {\bibfnamefont {T.}~\bibnamefont {Prokscha}}, \bibinfo {author}
  {\bibfnamefont {A.}~\bibnamefont {Suter}}, \bibinfo {author} {\bibfnamefont
  {Z.}~\bibnamefont {Salman}}, \bibinfo {author} {\bibfnamefont
  {M.}~\bibnamefont {Viret}}, \bibinfo {author} {\bibfnamefont
  {D.}~\bibnamefont {van~der Marel}}, \ and\ \bibinfo {author} {\bibfnamefont
  {S.}~\bibnamefont {Gariglio}},\ }\href {\doibase 10.1103/PhysRevB.103.125112}
  {\bibfield  {journal} {\bibinfo  {journal} {Phys. Rev. B}\ }\textbf {\bibinfo
  {volume} {103}},\ \bibinfo {pages} {125112} (\bibinfo {year}
  {2021})}\BibitemShut {NoStop}%
\bibitem [{\citenamefont {Mott}(1990)}]{Mott1990}%
  \BibitemOpen
  \bibfield  {author} {\bibinfo {author} {\bibfnamefont {N.~F.}\ \bibnamefont
  {Mott}},\ }\href@noop {} {\emph {\bibinfo {title} {Metal-Insulator
  Transitions}}},\ \bibinfo {edition} {2nd}\ ed.\ (\bibinfo  {publisher}
  {Taylor \& Francis},\ \bibinfo {address} {London},\ \bibinfo {year}
  {1990})\BibitemShut {NoStop}%
\bibitem [{\citenamefont {Imada}\ \emph {et~al.}(1998)\citenamefont {Imada},
  \citenamefont {Fujimori},\ and\ \citenamefont {Tokura}}]{Imada1998}%
  \BibitemOpen
  \bibfield  {author} {\bibinfo {author} {\bibfnamefont {M.}~\bibnamefont
  {Imada}}, \bibinfo {author} {\bibfnamefont {A.}~\bibnamefont {Fujimori}}, \
  and\ \bibinfo {author} {\bibfnamefont {Y.}~\bibnamefont {Tokura}},\ }\href
  {\doibase 10.1103/RevModPhys.70.1039} {\bibfield  {journal} {\bibinfo
  {journal} {Rev. Mod. Phys.}\ }\textbf {\bibinfo {volume} {70}},\ \bibinfo
  {pages} {1039} (\bibinfo {year} {1998})}\BibitemShut {NoStop}%
\bibitem [{\citenamefont {Fujimori}\ \emph {et~al.}(2001)\citenamefont
  {Fujimori}, \citenamefont {Yoshida}, \citenamefont {Okazaki}, \citenamefont
  {Tsujioka}, \citenamefont {Kobayashi}, \citenamefont {Mizokawa},
  \citenamefont {Onoda}, \citenamefont {Katsufuji}, \citenamefont {Taguchi},\
  and\ \citenamefont {Tokura}}]{FUJIMORI2001}%
  \BibitemOpen
  \bibfield  {author} {\bibinfo {author} {\bibfnamefont {A.}~\bibnamefont
  {Fujimori}}, \bibinfo {author} {\bibfnamefont {T.}~\bibnamefont {Yoshida}},
  \bibinfo {author} {\bibfnamefont {K.}~\bibnamefont {Okazaki}}, \bibinfo
  {author} {\bibfnamefont {T.}~\bibnamefont {Tsujioka}}, \bibinfo {author}
  {\bibfnamefont {K.}~\bibnamefont {Kobayashi}}, \bibinfo {author}
  {\bibfnamefont {T.}~\bibnamefont {Mizokawa}}, \bibinfo {author}
  {\bibfnamefont {M.}~\bibnamefont {Onoda}}, \bibinfo {author} {\bibfnamefont
  {T.}~\bibnamefont {Katsufuji}}, \bibinfo {author} {\bibfnamefont
  {Y.}~\bibnamefont {Taguchi}}, \ and\ \bibinfo {author} {\bibfnamefont
  {Y.}~\bibnamefont {Tokura}},\ }\href {\doibase
  https://doi.org/10.1016/S0368-2048(01)00253-5} {\bibfield  {journal}
  {\bibinfo  {journal} {J. Electron Spectrosc. Relat. Phenom.}\ }\textbf
  {\bibinfo {volume} {117-118}},\ \bibinfo {pages} {277 } (\bibinfo {year}
  {2001})}\BibitemShut {NoStop}%
\bibitem [{\citenamefont {Maiti}\ and\ \citenamefont
  {Sarma}(2000)}]{Maiti2000}%
  \BibitemOpen
  \bibfield  {author} {\bibinfo {author} {\bibfnamefont {K.}~\bibnamefont
  {Maiti}}\ and\ \bibinfo {author} {\bibfnamefont {D.~D.}\ \bibnamefont
  {Sarma}},\ }\href {\doibase 10.1103/PhysRevB.61.2525} {\bibfield  {journal}
  {\bibinfo  {journal} {Phys. Rev. B}\ }\textbf {\bibinfo {volume} {61}},\
  \bibinfo {pages} {2525} (\bibinfo {year} {2000})}\BibitemShut {NoStop}%
\bibitem [{\citenamefont {Zaanen}\ \emph {et~al.}(1985)\citenamefont {Zaanen},
  \citenamefont {Sawatzky},\ and\ \citenamefont {Allen}}]{Zaanen1985}%
  \BibitemOpen
  \bibfield  {author} {\bibinfo {author} {\bibfnamefont {J.}~\bibnamefont
  {Zaanen}}, \bibinfo {author} {\bibfnamefont {G.~A.}\ \bibnamefont
  {Sawatzky}}, \ and\ \bibinfo {author} {\bibfnamefont {J.~W.}\ \bibnamefont
  {Allen}},\ }\href {\doibase 10.1103/PhysRevLett.55.418} {\bibfield  {journal}
  {\bibinfo  {journal} {Phys. Rev. Lett.}\ }\textbf {\bibinfo {volume} {55}},\
  \bibinfo {pages} {418} (\bibinfo {year} {1985})}\BibitemShut {NoStop}%
\bibitem [{\citenamefont {Mahajan}\ \emph {et~al.}(1992)\citenamefont
  {Mahajan}, \citenamefont {Johnston}, \citenamefont {Torgeson},\ and\
  \citenamefont {Borsa}}]{Mahajan1992}%
  \BibitemOpen
  \bibfield  {author} {\bibinfo {author} {\bibfnamefont {A.~V.}\ \bibnamefont
  {Mahajan}}, \bibinfo {author} {\bibfnamefont {D.~C.}\ \bibnamefont
  {Johnston}}, \bibinfo {author} {\bibfnamefont {D.~R.}\ \bibnamefont
  {Torgeson}}, \ and\ \bibinfo {author} {\bibfnamefont {F.}~\bibnamefont
  {Borsa}},\ }\href {\doibase 10.1103/PhysRevB.46.10973} {\bibfield  {journal}
  {\bibinfo  {journal} {Phys. Rev. B}\ }\textbf {\bibinfo {volume} {46}},\
  \bibinfo {pages} {10973} (\bibinfo {year} {1992})}\BibitemShut {NoStop}%
\bibitem [{\citenamefont {Fujioka}\ \emph {et~al.}(2006)\citenamefont
  {Fujioka}, \citenamefont {Miyasaka},\ and\ \citenamefont
  {Tokura}}]{Fujioka2006}%
  \BibitemOpen
  \bibfield  {author} {\bibinfo {author} {\bibfnamefont {J.}~\bibnamefont
  {Fujioka}}, \bibinfo {author} {\bibfnamefont {S.}~\bibnamefont {Miyasaka}}, \
  and\ \bibinfo {author} {\bibfnamefont {Y.}~\bibnamefont {Tokura}},\ }\href
  {\doibase 10.1103/PhysRevLett.97.196401} {\bibfield  {journal} {\bibinfo
  {journal} {Phys. Rev. Lett.}\ }\textbf {\bibinfo {volume} {97}},\ \bibinfo
  {pages} {196401} (\bibinfo {year} {2006})}\BibitemShut {NoStop}%
\bibitem [{\citenamefont {Egdell}\ \emph {et~al.}(1984)\citenamefont {Egdell},
  \citenamefont {Harrison}, \citenamefont {Hill}, \citenamefont {Porte},\ and\
  \citenamefont {Wall}}]{Egdell1984}%
  \BibitemOpen
  \bibfield  {author} {\bibinfo {author} {\bibfnamefont {R.}~\bibnamefont
  {Egdell}}, \bibinfo {author} {\bibfnamefont {M.}~\bibnamefont {Harrison}},
  \bibinfo {author} {\bibfnamefont {M.}~\bibnamefont {Hill}}, \bibinfo {author}
  {\bibfnamefont {L.}~\bibnamefont {Porte}}, \ and\ \bibinfo {author}
  {\bibfnamefont {G.}~\bibnamefont {Wall}},\ }\href {\doibase
  10.1088/0022-3719/17/16/008} {\bibfield  {journal} {\bibinfo  {journal} {J.
  Phys. C}\ }\textbf {\bibinfo {volume} {17}},\ \bibinfo {pages} {2889}
  (\bibinfo {year} {1984})}\BibitemShut {NoStop}%
\bibitem [{\citenamefont {Miyasaka}\ \emph {et~al.}(2002)\citenamefont
  {Miyasaka}, \citenamefont {Okimoto},\ and\ \citenamefont
  {Tokura}}]{Miyasaka2002}%
  \BibitemOpen
  \bibfield  {author} {\bibinfo {author} {\bibfnamefont {S.}~\bibnamefont
  {Miyasaka}}, \bibinfo {author} {\bibfnamefont {Y.}~\bibnamefont {Okimoto}}, \
  and\ \bibinfo {author} {\bibfnamefont {Y.}~\bibnamefont {Tokura}},\ }\href
  {\doibase 10.1143/JPSJ.71.2086} {\bibfield  {journal} {\bibinfo  {journal}
  {J. Phys. Soc. Jpn}\ }\textbf {\bibinfo {volume} {71}},\ \bibinfo {pages}
  {2086} (\bibinfo {year} {2002})}\BibitemShut {NoStop}%
\bibitem [{\citenamefont {Fang}\ \emph {et~al.}(2003)\citenamefont {Fang},
  \citenamefont {Nagaosa},\ and\ \citenamefont {Terakura}}]{Fang2003}%
  \BibitemOpen
  \bibfield  {author} {\bibinfo {author} {\bibfnamefont {Z.}~\bibnamefont
  {Fang}}, \bibinfo {author} {\bibfnamefont {N.}~\bibnamefont {Nagaosa}}, \
  and\ \bibinfo {author} {\bibfnamefont {K.}~\bibnamefont {Terakura}},\ }\href
  {\doibase 10.1103/PhysRevB.67.035101} {\bibfield  {journal} {\bibinfo
  {journal} {Phys. Rev. B}\ }\textbf {\bibinfo {volume} {67}},\ \bibinfo
  {pages} {035101} (\bibinfo {year} {2003})}\BibitemShut {NoStop}%
\bibitem [{\citenamefont {Kim}(2018)}]{Kim2018}%
  \BibitemOpen
  \bibfield  {author} {\bibinfo {author} {\bibfnamefont {M.}~\bibnamefont
  {Kim}},\ }\href {\doibase 10.1103/PhysRevB.97.155141} {\bibfield  {journal}
  {\bibinfo  {journal} {Phys. Rev. B}\ }\textbf {\bibinfo {volume} {97}},\
  \bibinfo {pages} {155141} (\bibinfo {year} {2018})}\BibitemShut {NoStop}%
\bibitem [{\citenamefont {Ament}\ \emph
  {et~al.}(2011{\natexlab{a}})\citenamefont {Ament}, \citenamefont {van
  Veenendaal}, \citenamefont {Devereaux}, \citenamefont {Hill},\ and\
  \citenamefont {van~den Brink}}]{AMENT2011}%
  \BibitemOpen
  \bibfield  {author} {\bibinfo {author} {\bibfnamefont {L.~J.~P.}\
  \bibnamefont {Ament}}, \bibinfo {author} {\bibfnamefont {M.}~\bibnamefont
  {van Veenendaal}}, \bibinfo {author} {\bibfnamefont {T.~P.}\ \bibnamefont
  {Devereaux}}, \bibinfo {author} {\bibfnamefont {J.~P.}\ \bibnamefont {Hill}},
  \ and\ \bibinfo {author} {\bibfnamefont {J.}~\bibnamefont {van~den Brink}},\
  }\href {\doibase 10.1103/RevModPhys.83.705} {\bibfield  {journal} {\bibinfo
  {journal} {Rev. Mod. Phys.}\ }\textbf {\bibinfo {volume} {83}},\ \bibinfo
  {pages} {705} (\bibinfo {year} {2011}{\natexlab{a}})}\BibitemShut {NoStop}%
\bibitem [{\citenamefont {Brookes}\ \emph {et~al.}(2018)\citenamefont
  {Brookes}, \citenamefont {Yakhou-Harris}, \citenamefont {Kummer},
  \citenamefont {Fondacaro}, \citenamefont {Cezar}, \citenamefont {Betto},
  \citenamefont {Velez-Fort}, \citenamefont {Amorese}, \citenamefont
  {Ghiringhelli}, \citenamefont {Braicovich}, \citenamefont {Barrett},
  \citenamefont {Berruyer}, \citenamefont {Cianciosi}, \citenamefont {Eybert},
  \citenamefont {Marion}, \citenamefont {van~der Linden},\ and\ \citenamefont
  {Zhang}}]{BROOKES2018}%
  \BibitemOpen
  \bibfield  {author} {\bibinfo {author} {\bibfnamefont {N.}~\bibnamefont
  {Brookes}}, \bibinfo {author} {\bibfnamefont {F.}~\bibnamefont
  {Yakhou-Harris}}, \bibinfo {author} {\bibfnamefont {K.}~\bibnamefont
  {Kummer}}, \bibinfo {author} {\bibfnamefont {A.}~\bibnamefont {Fondacaro}},
  \bibinfo {author} {\bibfnamefont {J.}~\bibnamefont {Cezar}}, \bibinfo
  {author} {\bibfnamefont {D.}~\bibnamefont {Betto}}, \bibinfo {author}
  {\bibfnamefont {E.}~\bibnamefont {Velez-Fort}}, \bibinfo {author}
  {\bibfnamefont {A.}~\bibnamefont {Amorese}}, \bibinfo {author} {\bibfnamefont
  {G.}~\bibnamefont {Ghiringhelli}}, \bibinfo {author} {\bibfnamefont
  {L.}~\bibnamefont {Braicovich}}, \bibinfo {author} {\bibfnamefont
  {R.}~\bibnamefont {Barrett}}, \bibinfo {author} {\bibfnamefont
  {G.}~\bibnamefont {Berruyer}}, \bibinfo {author} {\bibfnamefont
  {F.}~\bibnamefont {Cianciosi}}, \bibinfo {author} {\bibfnamefont
  {L.}~\bibnamefont {Eybert}}, \bibinfo {author} {\bibfnamefont
  {P.}~\bibnamefont {Marion}}, \bibinfo {author} {\bibfnamefont
  {P.}~\bibnamefont {van~der Linden}}, \ and\ \bibinfo {author} {\bibfnamefont
  {L.}~\bibnamefont {Zhang}},\ }\href {\doibase
  https://doi.org/10.1016/j.nima.2018.07.001} {\bibfield  {journal} {\bibinfo
  {journal} {Nucl. Instrum. \& Methods A}\ }\textbf {\bibinfo {volume} {903}},\
  \bibinfo {pages} {175 } (\bibinfo {year} {2018})}\BibitemShut {NoStop}%
\bibitem [{\citenamefont {Amorese}\ \emph {et~al.}(2018)\citenamefont
  {Amorese}, \citenamefont {Caroca-Canales}, \citenamefont {Seiro},
  \citenamefont {Krellner}, \citenamefont {Ghiringhelli}, \citenamefont
  {Brookes}, \citenamefont {Vyalikh}, \citenamefont {Geibel},\ and\
  \citenamefont {Kummer}}]{AMORESE2018}%
  \BibitemOpen
  \bibfield  {author} {\bibinfo {author} {\bibfnamefont {A.}~\bibnamefont
  {Amorese}}, \bibinfo {author} {\bibfnamefont {N.}~\bibnamefont
  {Caroca-Canales}}, \bibinfo {author} {\bibfnamefont {S.}~\bibnamefont
  {Seiro}}, \bibinfo {author} {\bibfnamefont {C.}~\bibnamefont {Krellner}},
  \bibinfo {author} {\bibfnamefont {G.}~\bibnamefont {Ghiringhelli}}, \bibinfo
  {author} {\bibfnamefont {N.~B.}\ \bibnamefont {Brookes}}, \bibinfo {author}
  {\bibfnamefont {D.~V.}\ \bibnamefont {Vyalikh}}, \bibinfo {author}
  {\bibfnamefont {C.}~\bibnamefont {Geibel}}, \ and\ \bibinfo {author}
  {\bibfnamefont {K.}~\bibnamefont {Kummer}},\ }\href {\doibase
  10.1103/PhysRevB.97.245130} {\bibfield  {journal} {\bibinfo  {journal} {Phys.
  Rev. B}\ }\textbf {\bibinfo {volume} {97}},\ \bibinfo {pages} {245130}
  (\bibinfo {year} {2018})}\BibitemShut {NoStop}%
\bibitem [{\citenamefont {Amorese}\ \emph {et~al.}(2019)\citenamefont
  {Amorese}, \citenamefont {Stockert}, \citenamefont {Kummer}, \citenamefont
  {Brookes}, \citenamefont {Kim}, \citenamefont {Fisk}, \citenamefont
  {Haverkort}, \citenamefont {Thalmeier}, \citenamefont {Tjeng},\ and\
  \citenamefont {Severing}}]{AMORESE2019}%
  \BibitemOpen
  \bibfield  {author} {\bibinfo {author} {\bibfnamefont {A.}~\bibnamefont
  {Amorese}}, \bibinfo {author} {\bibfnamefont {O.}~\bibnamefont {Stockert}},
  \bibinfo {author} {\bibfnamefont {K.}~\bibnamefont {Kummer}}, \bibinfo
  {author} {\bibfnamefont {N.~B.}\ \bibnamefont {Brookes}}, \bibinfo {author}
  {\bibfnamefont {D.-J.}\ \bibnamefont {Kim}}, \bibinfo {author} {\bibfnamefont
  {Z.}~\bibnamefont {Fisk}}, \bibinfo {author} {\bibfnamefont {M.~W.}\
  \bibnamefont {Haverkort}}, \bibinfo {author} {\bibfnamefont {P.}~\bibnamefont
  {Thalmeier}}, \bibinfo {author} {\bibfnamefont {L.~H.}\ \bibnamefont
  {Tjeng}}, \ and\ \bibinfo {author} {\bibfnamefont {A.}~\bibnamefont
  {Severing}},\ }\href {\doibase 10.1103/PhysRevB.100.241107} {\bibfield
  {journal} {\bibinfo  {journal} {Phys. Rev. B}\ }\textbf {\bibinfo {volume}
  {100}},\ \bibinfo {pages} {241107(R)} (\bibinfo {year} {2019})}\BibitemShut
  {NoStop}%
\bibitem [{\citenamefont {Chen}\ \emph {et~al.}(2015)\citenamefont {Chen},
  \citenamefont {Laverock}, \citenamefont {Jr}, \citenamefont {McNulty},
  \citenamefont {Smith}, \citenamefont {Glans}, \citenamefont {Guo},
  \citenamefont {Qiao}, \citenamefont {Yang}, \citenamefont {Lees},
  \citenamefont {Tung}, \citenamefont {Singh},\ and\ \citenamefont
  {Balakrishnan}}]{CHEN2015}%
  \BibitemOpen
  \bibfield  {author} {\bibinfo {author} {\bibfnamefont {B.}~\bibnamefont
  {Chen}}, \bibinfo {author} {\bibfnamefont {J.}~\bibnamefont {Laverock}},
  \bibinfo {author} {\bibfnamefont {D.~N.}\ \bibnamefont {Jr}}, \bibinfo
  {author} {\bibfnamefont {J.~F.}\ \bibnamefont {McNulty}}, \bibinfo {author}
  {\bibfnamefont {K.~E.}\ \bibnamefont {Smith}}, \bibinfo {author}
  {\bibfnamefont {P.-A.}\ \bibnamefont {Glans}}, \bibinfo {author}
  {\bibfnamefont {J.-H.}\ \bibnamefont {Guo}}, \bibinfo {author} {\bibfnamefont
  {R.-M.}\ \bibnamefont {Qiao}}, \bibinfo {author} {\bibfnamefont {W.-L.}\
  \bibnamefont {Yang}}, \bibinfo {author} {\bibfnamefont {M.~R.}\ \bibnamefont
  {Lees}}, \bibinfo {author} {\bibfnamefont {L.~D.}\ \bibnamefont {Tung}},
  \bibinfo {author} {\bibfnamefont {R.~P.}\ \bibnamefont {Singh}}, \ and\
  \bibinfo {author} {\bibfnamefont {G.}~\bibnamefont {Balakrishnan}},\ }\href
  {http://stacks.iop.org/0953-8984/27/i=10/a=105503} {\bibfield  {journal}
  {\bibinfo  {journal} {J. Phys.: Condens. Matter}\ }\textbf {\bibinfo {volume}
  {27}},\ \bibinfo {pages} {105503} (\bibinfo {year} {2015})}\BibitemShut
  {NoStop}%
\bibitem [{\citenamefont {Laverock}\ \emph {et~al.}(2014)\citenamefont
  {Laverock}, \citenamefont {Chen}, \citenamefont {Preston}, \citenamefont
  {Newby}, \citenamefont {Piper}, \citenamefont {Tung}, \citenamefont
  {Balakrishnan}, \citenamefont {Glans}, \citenamefont {Guo},\ and\
  \citenamefont {Smith}}]{LAVEROCK2014}%
  \BibitemOpen
  \bibfield  {author} {\bibinfo {author} {\bibfnamefont {J.}~\bibnamefont
  {Laverock}}, \bibinfo {author} {\bibfnamefont {B.}~\bibnamefont {Chen}},
  \bibinfo {author} {\bibfnamefont {A.~R.~H.}\ \bibnamefont {Preston}},
  \bibinfo {author} {\bibfnamefont {D.}~\bibnamefont {Newby}}, \bibinfo
  {author} {\bibfnamefont {L.~F.~J.}\ \bibnamefont {Piper}}, \bibinfo {author}
  {\bibfnamefont {L.~D.}\ \bibnamefont {Tung}}, \bibinfo {author}
  {\bibfnamefont {G.}~\bibnamefont {Balakrishnan}}, \bibinfo {author}
  {\bibfnamefont {P.-A.}\ \bibnamefont {Glans}}, \bibinfo {author}
  {\bibfnamefont {J.-H.}\ \bibnamefont {Guo}}, \ and\ \bibinfo {author}
  {\bibfnamefont {K.~E.}\ \bibnamefont {Smith}},\ }\href@noop {} {\bibfield
  {journal} {\bibinfo  {journal} {J. Phys.: Condens. Matter}\ }\textbf
  {\bibinfo {volume} {26}},\ \bibinfo {pages} {455603} (\bibinfo {year}
  {2014})}\BibitemShut {NoStop}%
\bibitem [{\citenamefont {Ament}\ \emph
  {et~al.}(2011{\natexlab{b}})\citenamefont {Ament}, \citenamefont {van
  Veenendaal},\ and\ \citenamefont {van~den Brink}}]{Ament2011b}%
  \BibitemOpen
  \bibfield  {author} {\bibinfo {author} {\bibfnamefont {L.~J.~P.}\
  \bibnamefont {Ament}}, \bibinfo {author} {\bibfnamefont {M.}~\bibnamefont
  {van Veenendaal}}, \ and\ \bibinfo {author} {\bibfnamefont {J.}~\bibnamefont
  {van~den Brink}},\ }\href {\doibase 10.1209/0295-5075/95/27008} {\bibfield
  {journal} {\bibinfo  {journal} {{EPL} (Europhysics Letters)}\ }\textbf
  {\bibinfo {volume} {95}},\ \bibinfo {pages} {27008} (\bibinfo {year}
  {2011}{\natexlab{b}})}\BibitemShut {NoStop}%
\bibitem [{\citenamefont {Sacchi}\ \emph {et~al.}(2013)\citenamefont {Sacchi},
  \citenamefont {Jaouen}, \citenamefont {Popescu}, \citenamefont {Gaudemer},
  \citenamefont {Tonnerre}, \citenamefont {Chiuzbaian}, \citenamefont {Hague},
  \citenamefont {Delmotte}, \citenamefont {Dubuisson}, \citenamefont {Cauchon},
  \citenamefont {Lagarde},\ and\ \citenamefont {Polack}}]{SACCHI2013}%
  \BibitemOpen
  \bibfield  {author} {\bibinfo {author} {\bibfnamefont {M.}~\bibnamefont
  {Sacchi}}, \bibinfo {author} {\bibfnamefont {N.}~\bibnamefont {Jaouen}},
  \bibinfo {author} {\bibfnamefont {H.}~\bibnamefont {Popescu}}, \bibinfo
  {author} {\bibfnamefont {R.}~\bibnamefont {Gaudemer}}, \bibinfo {author}
  {\bibfnamefont {J.~M.}\ \bibnamefont {Tonnerre}}, \bibinfo {author}
  {\bibfnamefont {S.~G.}\ \bibnamefont {Chiuzbaian}}, \bibinfo {author}
  {\bibfnamefont {C.~F.}\ \bibnamefont {Hague}}, \bibinfo {author}
  {\bibfnamefont {A.}~\bibnamefont {Delmotte}}, \bibinfo {author}
  {\bibfnamefont {J.~M.}\ \bibnamefont {Dubuisson}}, \bibinfo {author}
  {\bibfnamefont {G.}~\bibnamefont {Cauchon}}, \bibinfo {author} {\bibfnamefont
  {B.}~\bibnamefont {Lagarde}}, \ and\ \bibinfo {author} {\bibfnamefont
  {F.}~\bibnamefont {Polack}},\ }\href
  {http://stacks.iop.org/1742-6596/425/i=7/a=072018} {\bibfield  {journal}
  {\bibinfo  {journal} {J. Phys. Conf. Ser.}\ }\textbf {\bibinfo {volume}
  {425}},\ \bibinfo {pages} {072018} (\bibinfo {year} {2013})}\BibitemShut
  {NoStop}%
\bibitem [{\citenamefont {Chiuzb{\u{a}}ian}\ \emph {et~al.}(2014)\citenamefont
  {Chiuzb{\u{a}}ian}, \citenamefont {Hague}, \citenamefont {Avila},
  \citenamefont {Delaunay}, \citenamefont {Jaouen}, \citenamefont {Sacchi},
  \citenamefont {Polack}, \citenamefont {Thomasset}, \citenamefont {Lagarde},
  \citenamefont {Nicolaou} \emph {et~al.}}]{CHIUZBUAIAN2014}%
  \BibitemOpen
  \bibfield  {author} {\bibinfo {author} {\bibfnamefont {S.~G.}\ \bibnamefont
  {Chiuzb{\u{a}}ian}}, \bibinfo {author} {\bibfnamefont {C.~F.}\ \bibnamefont
  {Hague}}, \bibinfo {author} {\bibfnamefont {A.}~\bibnamefont {Avila}},
  \bibinfo {author} {\bibfnamefont {R.}~\bibnamefont {Delaunay}}, \bibinfo
  {author} {\bibfnamefont {N.}~\bibnamefont {Jaouen}}, \bibinfo {author}
  {\bibfnamefont {M.}~\bibnamefont {Sacchi}}, \bibinfo {author} {\bibfnamefont
  {F.}~\bibnamefont {Polack}}, \bibinfo {author} {\bibfnamefont
  {M.}~\bibnamefont {Thomasset}}, \bibinfo {author} {\bibfnamefont
  {B.}~\bibnamefont {Lagarde}}, \bibinfo {author} {\bibfnamefont
  {A.}~\bibnamefont {Nicolaou}},  \emph {et~al.},\ }\href@noop {} {\bibfield
  {journal} {\bibinfo  {journal} {Rev, of Sci. Instrum.}\ }\textbf {\bibinfo
  {volume} {85}},\ \bibinfo {pages} {043108} (\bibinfo {year}
  {2014})}\BibitemShut {NoStop}%
\bibitem [{\citenamefont {Zhang}\ \emph {et~al.}(2017)\citenamefont {Zhang},
  \citenamefont {Brahlek}, \citenamefont {Ji}, \citenamefont {Lei},
  \citenamefont {Lapano}, \citenamefont {Freeland}, \citenamefont {Gopalan},\
  and\ \citenamefont {Engel-Herbert}}]{ZHANG2017}%
  \BibitemOpen
  \bibfield  {author} {\bibinfo {author} {\bibfnamefont {H.-T.}\ \bibnamefont
  {Zhang}}, \bibinfo {author} {\bibfnamefont {M.}~\bibnamefont {Brahlek}},
  \bibinfo {author} {\bibfnamefont {X.}~\bibnamefont {Ji}}, \bibinfo {author}
  {\bibfnamefont {S.}~\bibnamefont {Lei}}, \bibinfo {author} {\bibfnamefont
  {J.}~\bibnamefont {Lapano}}, \bibinfo {author} {\bibfnamefont {J.~W.}\
  \bibnamefont {Freeland}}, \bibinfo {author} {\bibfnamefont {V.}~\bibnamefont
  {Gopalan}}, \ and\ \bibinfo {author} {\bibfnamefont {R.}~\bibnamefont
  {Engel-Herbert}},\ }\href {\doibase 10.1021/acsami.6b16007} {\bibfield
  {journal} {\bibinfo  {journal} {ACS Appl. Mater. Interfaces}\ }\textbf
  {\bibinfo {volume} {9}},\ \bibinfo {pages} {12556} (\bibinfo {year}
  {2017})}\BibitemShut {NoStop}%
\bibitem [{\citenamefont {Wadati}\ \emph {et~al.}(2009)\citenamefont {Wadati},
  \citenamefont {Hawthorn}, \citenamefont {Geck}, \citenamefont {Regier},
  \citenamefont {Blyth}, \citenamefont {Higuchi}, \citenamefont {Hotta},
  \citenamefont {Hikita}, \citenamefont {Hwang},\ and\ \citenamefont
  {Sawatzky}}]{WADATI2009}%
  \BibitemOpen
  \bibfield  {author} {\bibinfo {author} {\bibfnamefont {H.}~\bibnamefont
  {Wadati}}, \bibinfo {author} {\bibfnamefont {D.~G.}\ \bibnamefont
  {Hawthorn}}, \bibinfo {author} {\bibfnamefont {J.}~\bibnamefont {Geck}},
  \bibinfo {author} {\bibfnamefont {T.~Z.}\ \bibnamefont {Regier}}, \bibinfo
  {author} {\bibfnamefont {R.~I.~R.}\ \bibnamefont {Blyth}}, \bibinfo {author}
  {\bibfnamefont {T.}~\bibnamefont {Higuchi}}, \bibinfo {author} {\bibfnamefont
  {Y.}~\bibnamefont {Hotta}}, \bibinfo {author} {\bibfnamefont
  {Y.}~\bibnamefont {Hikita}}, \bibinfo {author} {\bibfnamefont {H.~Y.}\
  \bibnamefont {Hwang}}, \ and\ \bibinfo {author} {\bibfnamefont {G.~A.}\
  \bibnamefont {Sawatzky}},\ }\href {\doibase 10.1063/1.3177328} {\bibfield
  {journal} {\bibinfo  {journal} {Appl. Phys. Lett.}\ }\textbf {\bibinfo
  {volume} {95}},\ \bibinfo {pages} {023115} (\bibinfo {year} {2009})},\
  \Eprint {http://arxiv.org/abs/https://doi.org/10.1063/1.3177328}
  {https://doi.org/10.1063/1.3177328} \BibitemShut {NoStop}%
\bibitem [{\citenamefont {Pen}\ \emph {et~al.}(1999)\citenamefont {Pen},
  \citenamefont {Abbate}, \citenamefont {Fuijmori}, \citenamefont {Tokura},
  \citenamefont {Eisaki}, \citenamefont {Uchida},\ and\ \citenamefont
  {Sawatzky}}]{PEN1999}%
  \BibitemOpen
  \bibfield  {author} {\bibinfo {author} {\bibfnamefont {H.~F.}\ \bibnamefont
  {Pen}}, \bibinfo {author} {\bibfnamefont {M.}~\bibnamefont {Abbate}},
  \bibinfo {author} {\bibfnamefont {A.}~\bibnamefont {Fuijmori}}, \bibinfo
  {author} {\bibfnamefont {Y.}~\bibnamefont {Tokura}}, \bibinfo {author}
  {\bibfnamefont {H.}~\bibnamefont {Eisaki}}, \bibinfo {author} {\bibfnamefont
  {S.}~\bibnamefont {Uchida}}, \ and\ \bibinfo {author} {\bibfnamefont {G.~A.}\
  \bibnamefont {Sawatzky}},\ }\href {\doibase 10.1103/PhysRevB.59.7422}
  {\bibfield  {journal} {\bibinfo  {journal} {Phys. Rev. B}\ }\textbf {\bibinfo
  {volume} {59}},\ \bibinfo {pages} {7422} (\bibinfo {year}
  {1999})}\BibitemShut {NoStop}%
\bibitem [{\citenamefont {Park}\ \emph {et~al.}(2000)\citenamefont {Park},
  \citenamefont {Tjeng}, \citenamefont {Tanaka}, \citenamefont {Allen},
  \citenamefont {Chen}, \citenamefont {Metcalf}, \citenamefont {Honig},
  \citenamefont {de~Groot},\ and\ \citenamefont {Sawatzky}}]{PARK2000}%
  \BibitemOpen
  \bibfield  {author} {\bibinfo {author} {\bibfnamefont {J.-H.}\ \bibnamefont
  {Park}}, \bibinfo {author} {\bibfnamefont {L.~H.}\ \bibnamefont {Tjeng}},
  \bibinfo {author} {\bibfnamefont {A.}~\bibnamefont {Tanaka}}, \bibinfo
  {author} {\bibfnamefont {J.~W.}\ \bibnamefont {Allen}}, \bibinfo {author}
  {\bibfnamefont {C.~T.}\ \bibnamefont {Chen}}, \bibinfo {author}
  {\bibfnamefont {P.}~\bibnamefont {Metcalf}}, \bibinfo {author} {\bibfnamefont
  {J.~M.}\ \bibnamefont {Honig}}, \bibinfo {author} {\bibfnamefont {F.~M.~F.}\
  \bibnamefont {de~Groot}}, \ and\ \bibinfo {author} {\bibfnamefont {G.~A.}\
  \bibnamefont {Sawatzky}},\ }\href {\doibase 10.1103/PhysRevB.61.11506}
  {\bibfield  {journal} {\bibinfo  {journal} {Phys. Rev. B}\ }\textbf {\bibinfo
  {volume} {61}},\ \bibinfo {pages} {11506} (\bibinfo {year}
  {2000})}\BibitemShut {NoStop}%
\bibitem [{\citenamefont {Schmitt}\ \emph {et~al.}(2002)\citenamefont
  {Schmitt}, \citenamefont {Duda}, \citenamefont {Augustsson}, \citenamefont
  {Guo}, \citenamefont {Nordgren}, \citenamefont {Downes}, \citenamefont
  {McGuinness}, \citenamefont {Smith}, \citenamefont {Dhalenne}, \citenamefont
  {Revcolevschi}, \citenamefont {Klemm},\ and\ \citenamefont
  {Horn}}]{schmitt2002}%
  \BibitemOpen
  \bibfield  {author} {\bibinfo {author} {\bibfnamefont {T.}~\bibnamefont
  {Schmitt}}, \bibinfo {author} {\bibfnamefont {L.-C.}\ \bibnamefont {Duda}},
  \bibinfo {author} {\bibfnamefont {A.}~\bibnamefont {Augustsson}}, \bibinfo
  {author} {\bibfnamefont {J.-H.}\ \bibnamefont {Guo}}, \bibinfo {author}
  {\bibfnamefont {J.}~\bibnamefont {Nordgren}}, \bibinfo {author}
  {\bibfnamefont {J.~E.}\ \bibnamefont {Downes}}, \bibinfo {author}
  {\bibfnamefont {C.}~\bibnamefont {McGuinness}}, \bibinfo {author}
  {\bibfnamefont {K.~E.}\ \bibnamefont {Smith}}, \bibinfo {author}
  {\bibfnamefont {G.}~\bibnamefont {Dhalenne}}, \bibinfo {author}
  {\bibfnamefont {A.}~\bibnamefont {Revcolevschi}}, \bibinfo {author}
  {\bibfnamefont {M.}~\bibnamefont {Klemm}}, \ and\ \bibinfo {author}
  {\bibfnamefont {S.}~\bibnamefont {Horn}},\ }\href {\doibase
  10.1142/S0218625X02003822} {\bibfield  {journal} {\bibinfo  {journal} {Surf.
  Rev. Lett.}\ }\textbf {\bibinfo {volume} {09}},\ \bibinfo {pages} {1369}
  (\bibinfo {year} {2002})}\BibitemShut {NoStop}%
\bibitem [{\citenamefont {Haverkort}\ \emph {et~al.}(2005)\citenamefont
  {Haverkort}, \citenamefont {Hu}, \citenamefont {Tanaka}, \citenamefont
  {Reichelt}, \citenamefont {Streltsov}, \citenamefont {Korotin}, \citenamefont
  {Anisimov}, \citenamefont {Hsieh}, \citenamefont {Lin}, \citenamefont {Chen},
  \citenamefont {Khomskii},\ and\ \citenamefont {Tjeng}}]{HAVERKORT2005}%
  \BibitemOpen
  \bibfield  {author} {\bibinfo {author} {\bibfnamefont {M.~W.}\ \bibnamefont
  {Haverkort}}, \bibinfo {author} {\bibfnamefont {Z.}~\bibnamefont {Hu}},
  \bibinfo {author} {\bibfnamefont {A.}~\bibnamefont {Tanaka}}, \bibinfo
  {author} {\bibfnamefont {W.}~\bibnamefont {Reichelt}}, \bibinfo {author}
  {\bibfnamefont {S.~V.}\ \bibnamefont {Streltsov}}, \bibinfo {author}
  {\bibfnamefont {M.~A.}\ \bibnamefont {Korotin}}, \bibinfo {author}
  {\bibfnamefont {V.~I.}\ \bibnamefont {Anisimov}}, \bibinfo {author}
  {\bibfnamefont {H.~H.}\ \bibnamefont {Hsieh}}, \bibinfo {author}
  {\bibfnamefont {H.-J.}\ \bibnamefont {Lin}}, \bibinfo {author} {\bibfnamefont
  {C.~T.}\ \bibnamefont {Chen}}, \bibinfo {author} {\bibfnamefont {D.~I.}\
  \bibnamefont {Khomskii}}, \ and\ \bibinfo {author} {\bibfnamefont {L.~H.}\
  \bibnamefont {Tjeng}},\ }\href {\doibase 10.1103/PhysRevLett.95.196404}
  {\bibfield  {journal} {\bibinfo  {journal} {Phys. Rev. Lett.}\ }\textbf
  {\bibinfo {volume} {95}},\ \bibinfo {pages} {196404} (\bibinfo {year}
  {2005})}\BibitemShut {NoStop}%
\bibitem [{\citenamefont {Wadati}(2006)}]{WADATIthesis}%
  \BibitemOpen
  \bibfield  {author} {\bibinfo {author} {\bibfnamefont {H.}~\bibnamefont
  {Wadati}},\ }\href@noop {} {Ph.D. thesis},\ \bibinfo  {school} {University of
  Tokyo} (\bibinfo {year} {2006})\BibitemShut {NoStop}%
\bibitem [{\citenamefont {Mossanek}\ \emph {et~al.}(2009)\citenamefont
  {Mossanek}, \citenamefont {Abbate}, \citenamefont {Fonseca}, \citenamefont
  {Fujimori}, \citenamefont {Eisaki}, \citenamefont {Uchida},\ and\
  \citenamefont {Tokura}}]{MOSSANEK2009}%
  \BibitemOpen
  \bibfield  {author} {\bibinfo {author} {\bibfnamefont {R.~J.~O.}\
  \bibnamefont {Mossanek}}, \bibinfo {author} {\bibfnamefont {M.}~\bibnamefont
  {Abbate}}, \bibinfo {author} {\bibfnamefont {P.~T.}\ \bibnamefont {Fonseca}},
  \bibinfo {author} {\bibfnamefont {A.}~\bibnamefont {Fujimori}}, \bibinfo
  {author} {\bibfnamefont {H.}~\bibnamefont {Eisaki}}, \bibinfo {author}
  {\bibfnamefont {S.}~\bibnamefont {Uchida}}, \ and\ \bibinfo {author}
  {\bibfnamefont {Y.}~\bibnamefont {Tokura}},\ }\href {\doibase
  10.1103/PhysRevB.80.195107} {\bibfield  {journal} {\bibinfo  {journal} {Phys.
  Rev. B}\ }\textbf {\bibinfo {volume} {80}},\ \bibinfo {pages} {195107}
  (\bibinfo {year} {2009})}\BibitemShut {NoStop}%
\bibitem [{\citenamefont {Chen}\ \emph {et~al.}(1991)\citenamefont {Chen},
  \citenamefont {Sette}, \citenamefont {Ma}, \citenamefont {Hybertsen},
  \citenamefont {Stechel}, \citenamefont {Foulkes}, \citenamefont {Schluter},
  \citenamefont {Cheong}, \citenamefont {Cooper}, \citenamefont {Rupp},
  \citenamefont {Batlogg}, \citenamefont {Soo}, \citenamefont {Ming},
  \citenamefont {Krol},\ and\ \citenamefont {Kao}}]{PhysRevLett.66.104}%
  \BibitemOpen
  \bibfield  {author} {\bibinfo {author} {\bibfnamefont {C.~T.}\ \bibnamefont
  {Chen}}, \bibinfo {author} {\bibfnamefont {F.}~\bibnamefont {Sette}},
  \bibinfo {author} {\bibfnamefont {Y.}~\bibnamefont {Ma}}, \bibinfo {author}
  {\bibfnamefont {M.~S.}\ \bibnamefont {Hybertsen}}, \bibinfo {author}
  {\bibfnamefont {E.~B.}\ \bibnamefont {Stechel}}, \bibinfo {author}
  {\bibfnamefont {W.~M.~C.}\ \bibnamefont {Foulkes}}, \bibinfo {author}
  {\bibfnamefont {M.}~\bibnamefont {Schluter}}, \bibinfo {author}
  {\bibfnamefont {S.-W.}\ \bibnamefont {Cheong}}, \bibinfo {author}
  {\bibfnamefont {A.~S.}\ \bibnamefont {Cooper}}, \bibinfo {author}
  {\bibfnamefont {L.~W.}\ \bibnamefont {Rupp}}, \bibinfo {author}
  {\bibfnamefont {B.}~\bibnamefont {Batlogg}}, \bibinfo {author} {\bibfnamefont
  {Y.~L.}\ \bibnamefont {Soo}}, \bibinfo {author} {\bibfnamefont {Z.~H.}\
  \bibnamefont {Ming}}, \bibinfo {author} {\bibfnamefont {A.}~\bibnamefont
  {Krol}}, \ and\ \bibinfo {author} {\bibfnamefont {Y.~H.}\ \bibnamefont
  {Kao}},\ }\href {\doibase 10.1103/PhysRevLett.66.104} {\bibfield  {journal}
  {\bibinfo  {journal} {Phys. Rev. Lett.}\ }\textbf {\bibinfo {volume} {66}},\
  \bibinfo {pages} {104} (\bibinfo {year} {1991})}\BibitemShut {NoStop}%
\bibitem [{\citenamefont {van Elp}\ \emph {et~al.}(1992)\citenamefont {van
  Elp}, \citenamefont {Eskes}, \citenamefont {Kuiper},\ and\ \citenamefont
  {Sawatzky}}]{PhysRevB.45.1612}%
  \BibitemOpen
  \bibfield  {author} {\bibinfo {author} {\bibfnamefont {J.}~\bibnamefont {van
  Elp}}, \bibinfo {author} {\bibfnamefont {H.}~\bibnamefont {Eskes}}, \bibinfo
  {author} {\bibfnamefont {P.}~\bibnamefont {Kuiper}}, \ and\ \bibinfo {author}
  {\bibfnamefont {G.~A.}\ \bibnamefont {Sawatzky}},\ }\href {\doibase
  10.1103/PhysRevB.45.1612} {\bibfield  {journal} {\bibinfo  {journal} {Phys.
  Rev. B}\ }\textbf {\bibinfo {volume} {45}},\ \bibinfo {pages} {1612}
  (\bibinfo {year} {1992})}\BibitemShut {NoStop}%
\bibitem [{\citenamefont {Yushankhai}\ and\ \citenamefont
  {Siurakshina}(2013)}]{YUSHANKAI2013}%
  \BibitemOpen
  \bibfield  {author} {\bibinfo {author} {\bibfnamefont {V.}~\bibnamefont
  {Yushankhai}}\ and\ \bibinfo {author} {\bibfnamefont {L.}~\bibnamefont
  {Siurakshina}},\ }\href {\doibase 10.1142/S0217979213501853} {\bibfield
  {journal} {\bibinfo  {journal} {Int. J. Mod. Phys. B}\ }\textbf {\bibinfo
  {volume} {27}},\ \bibinfo {pages} {1350185} (\bibinfo {year}
  {2013})}\BibitemShut {NoStop}%
\bibitem [{\citenamefont {De~Groot}\ and\ \citenamefont
  {Kotani}(2008)}]{DEGROOT2008}%
  \BibitemOpen
  \bibfield  {author} {\bibinfo {author} {\bibfnamefont {F.}~\bibnamefont
  {De~Groot}}\ and\ \bibinfo {author} {\bibfnamefont {A.}~\bibnamefont
  {Kotani}},\ }\href@noop {} {\emph {\bibinfo {title} {Core Level Spectroscopy
  of Solids}}},\ Advances in Condensed Matter Science\ (\bibinfo  {publisher}
  {Taylor \& Francis Group},\ \bibinfo {year} {2008})\BibitemShut {NoStop}%
\bibitem [{\citenamefont {Bogdanov}\ \emph {et~al.}(2011)\citenamefont
  {Bogdanov}, \citenamefont {van~den Brink},\ and\ \citenamefont
  {Hozoi}}]{BOGDANOV2011}%
  \BibitemOpen
  \bibfield  {author} {\bibinfo {author} {\bibfnamefont {N.~A.}\ \bibnamefont
  {Bogdanov}}, \bibinfo {author} {\bibfnamefont {J.}~\bibnamefont {van~den
  Brink}}, \ and\ \bibinfo {author} {\bibfnamefont {L.}~\bibnamefont {Hozoi}},\
  }\href {\doibase 10.1103/PhysRevB.84.235146} {\bibfield  {journal} {\bibinfo
  {journal} {Phys. Rev. B}\ }\textbf {\bibinfo {volume} {84}},\ \bibinfo
  {pages} {235146} (\bibinfo {year} {2011})}\BibitemShut {NoStop}%
\bibitem [{\citenamefont {Ulrich}\ \emph {et~al.}(2008)\citenamefont {Ulrich},
  \citenamefont {Ghiringhelli}, \citenamefont {Piazzalunga}, \citenamefont
  {Braicovich}, \citenamefont {Brookes}, \citenamefont {Roth}, \citenamefont
  {Lorenz},\ and\ \citenamefont {Keimer}}]{ULRICH2008}%
  \BibitemOpen
  \bibfield  {author} {\bibinfo {author} {\bibfnamefont {C.}~\bibnamefont
  {Ulrich}}, \bibinfo {author} {\bibfnamefont {G.}~\bibnamefont
  {Ghiringhelli}}, \bibinfo {author} {\bibfnamefont {A.}~\bibnamefont
  {Piazzalunga}}, \bibinfo {author} {\bibfnamefont {L.}~\bibnamefont
  {Braicovich}}, \bibinfo {author} {\bibfnamefont {N.~B.}\ \bibnamefont
  {Brookes}}, \bibinfo {author} {\bibfnamefont {H.}~\bibnamefont {Roth}},
  \bibinfo {author} {\bibfnamefont {T.}~\bibnamefont {Lorenz}}, \ and\ \bibinfo
  {author} {\bibfnamefont {B.}~\bibnamefont {Keimer}},\ }\href {\doibase
  10.1103/PhysRevB.77.113102} {\bibfield  {journal} {\bibinfo  {journal} {Phys.
  Rev. B}\ }\textbf {\bibinfo {volume} {77}},\ \bibinfo {pages} {113102}
  (\bibinfo {year} {2008})}\BibitemShut {NoStop}%
\bibitem [{\citenamefont {Pavarini}\ \emph {et~al.}(2005)\citenamefont
  {Pavarini}, \citenamefont {Yamasaki}, \citenamefont {Nuss},\ and\
  \citenamefont {Andersen}}]{PAVARINI2005}%
  \BibitemOpen
  \bibfield  {author} {\bibinfo {author} {\bibfnamefont {E.}~\bibnamefont
  {Pavarini}}, \bibinfo {author} {\bibfnamefont {A.}~\bibnamefont {Yamasaki}},
  \bibinfo {author} {\bibfnamefont {J.}~\bibnamefont {Nuss}}, \ and\ \bibinfo
  {author} {\bibfnamefont {O.~K.}\ \bibnamefont {Andersen}},\ }\href {\doibase
  10.1088/1367-2630/7/1/188} {\bibfield  {journal} {\bibinfo  {journal} {New
  Journal of Physics}\ }\textbf {\bibinfo {volume} {7}},\ \bibinfo {pages}
  {188} (\bibinfo {year} {2005})}\BibitemShut {NoStop}%
\bibitem [{\citenamefont {Haule}\ \emph {et~al.}(2014)\citenamefont {Haule},
  \citenamefont {Birol},\ and\ \citenamefont {Kotliar}}]{HAULE2014}%
  \BibitemOpen
  \bibfield  {author} {\bibinfo {author} {\bibfnamefont {K.}~\bibnamefont
  {Haule}}, \bibinfo {author} {\bibfnamefont {T.}~\bibnamefont {Birol}}, \ and\
  \bibinfo {author} {\bibfnamefont {G.}~\bibnamefont {Kotliar}},\ }\href
  {\doibase 10.1103/PhysRevB.90.075136} {\bibfield  {journal} {\bibinfo
  {journal} {Phys. Rev. B}\ }\textbf {\bibinfo {volume} {90}},\ \bibinfo
  {pages} {075136} (\bibinfo {year} {2014})}\BibitemShut {NoStop}%
\bibitem [{\citenamefont {Geondzhian}\ and\ \citenamefont
  {Gilmore}(2018)}]{GEONDZHIAN2018}%
  \BibitemOpen
  \bibfield  {author} {\bibinfo {author} {\bibfnamefont {A.}~\bibnamefont
  {Geondzhian}}\ and\ \bibinfo {author} {\bibfnamefont {K.}~\bibnamefont
  {Gilmore}},\ }\href {\doibase 10.1103/PhysRevB.98.214305} {\bibfield
  {journal} {\bibinfo  {journal} {Phys. Rev. B}\ }\textbf {\bibinfo {volume}
  {98}},\ \bibinfo {pages} {214305} (\bibinfo {year} {2018})}\BibitemShut
  {NoStop}%
\bibitem [{\citenamefont {Geondzhian}\ and\ \citenamefont
  {Gilmore}(2020)}]{GEONDZHIAN2020}%
  \BibitemOpen
  \bibfield  {author} {\bibinfo {author} {\bibfnamefont {A.}~\bibnamefont
  {Geondzhian}}\ and\ \bibinfo {author} {\bibfnamefont {K.}~\bibnamefont
  {Gilmore}},\ }\href {\doibase 10.1103/PhysRevB.101.214307} {\bibfield
  {journal} {\bibinfo  {journal} {Phys. Rev. B}\ }\textbf {\bibinfo {volume}
  {101}},\ \bibinfo {pages} {214307} (\bibinfo {year} {2020})}\BibitemShut
  {NoStop}%
\bibitem [{\citenamefont {Choudhury}\ \emph {et~al.}(2008)\citenamefont
  {Choudhury}, \citenamefont {Walter}, \citenamefont {Kolesnikov},\ and\
  \citenamefont {Loong}}]{CHOUDHURY2008}%
  \BibitemOpen
  \bibfield  {author} {\bibinfo {author} {\bibfnamefont {N.}~\bibnamefont
  {Choudhury}}, \bibinfo {author} {\bibfnamefont {E.~J.}\ \bibnamefont
  {Walter}}, \bibinfo {author} {\bibfnamefont {A.~I.}\ \bibnamefont
  {Kolesnikov}}, \ and\ \bibinfo {author} {\bibfnamefont {C.-K.}\ \bibnamefont
  {Loong}},\ }\href {\doibase 10.1103/PhysRevB.77.134111} {\bibfield  {journal}
  {\bibinfo  {journal} {Phys. Rev. B}\ }\textbf {\bibinfo {volume} {77}},\
  \bibinfo {pages} {134111} (\bibinfo {year} {2008})}\BibitemShut {NoStop}%
\bibitem [{\citenamefont {Vaz~da Cruz}\ \emph {et~al.}(2019)\citenamefont
  {Vaz~da Cruz}, \citenamefont {Gel'mukhanov}, \citenamefont {Eckert},
  \citenamefont {Iannuzzi}, \citenamefont {Ertan}, \citenamefont {Pietzsch},
  \citenamefont {Couto}, \citenamefont {Niskanen}, \citenamefont {Fondell},
  \citenamefont {Dantz}, \citenamefont {Schmitt}, \citenamefont {Lu},
  \citenamefont {McNally}, \citenamefont {Jay}, \citenamefont {Kimberg},
  \citenamefont {F{\"o}hlisch},\ and\ \citenamefont {Odelius}}]{VazdaCruz2019}%
  \BibitemOpen
  \bibfield  {author} {\bibinfo {author} {\bibfnamefont {V.}~\bibnamefont
  {Vaz~da Cruz}}, \bibinfo {author} {\bibfnamefont {F.}~\bibnamefont
  {Gel'mukhanov}}, \bibinfo {author} {\bibfnamefont {S.}~\bibnamefont
  {Eckert}}, \bibinfo {author} {\bibfnamefont {M.}~\bibnamefont {Iannuzzi}},
  \bibinfo {author} {\bibfnamefont {E.}~\bibnamefont {Ertan}}, \bibinfo
  {author} {\bibfnamefont {A.}~\bibnamefont {Pietzsch}}, \bibinfo {author}
  {\bibfnamefont {R.~C.}\ \bibnamefont {Couto}}, \bibinfo {author}
  {\bibfnamefont {J.}~\bibnamefont {Niskanen}}, \bibinfo {author}
  {\bibfnamefont {M.}~\bibnamefont {Fondell}}, \bibinfo {author} {\bibfnamefont
  {M.}~\bibnamefont {Dantz}}, \bibinfo {author} {\bibfnamefont
  {T.}~\bibnamefont {Schmitt}}, \bibinfo {author} {\bibfnamefont
  {X.}~\bibnamefont {Lu}}, \bibinfo {author} {\bibfnamefont {D.}~\bibnamefont
  {McNally}}, \bibinfo {author} {\bibfnamefont {R.~M.}\ \bibnamefont {Jay}},
  \bibinfo {author} {\bibfnamefont {V.}~\bibnamefont {Kimberg}}, \bibinfo
  {author} {\bibfnamefont {A.}~\bibnamefont {F{\"o}hlisch}}, \ and\ \bibinfo
  {author} {\bibfnamefont {M.}~\bibnamefont {Odelius}},\ }\href {\doibase
  10.1038/s41467-019-08979-4} {\bibfield  {journal} {\bibinfo  {journal}
  {Nature Communications}\ }\textbf {\bibinfo {volume} {10}},\ \bibinfo {pages}
  {1013} (\bibinfo {year} {2019})}\BibitemShut {NoStop}%
\bibitem [{\citenamefont {Fujioka}\ \emph {et~al.}(2008)\citenamefont
  {Fujioka}, \citenamefont {Miyasaka},\ and\ \citenamefont
  {Tokura}}]{fujioka2008}%
  \BibitemOpen
  \bibfield  {author} {\bibinfo {author} {\bibfnamefont {J.}~\bibnamefont
  {Fujioka}}, \bibinfo {author} {\bibfnamefont {S.}~\bibnamefont {Miyasaka}}, \
  and\ \bibinfo {author} {\bibfnamefont {Y.}~\bibnamefont {Tokura}},\ }\href
  {\doibase 10.1103/PhysRevB.77.144402} {\bibfield  {journal} {\bibinfo
  {journal} {Phys. Rev. B}\ }\textbf {\bibinfo {volume} {77}},\ \bibinfo
  {pages} {144402} (\bibinfo {year} {2008})}\BibitemShut {NoStop}%
\bibitem [{\citenamefont {Sun}\ \emph {et~al.}(2011)\citenamefont {Sun},
  \citenamefont {Hennies}, \citenamefont {Pietzsch}, \citenamefont {Kennedy},
  \citenamefont {Schmitt}, \citenamefont {Strocov}, \citenamefont {Andersson},
  \citenamefont {Berglund}, \citenamefont {Rubensson}, \citenamefont {Aidas},
  \citenamefont {Gel'mukhanov}, \citenamefont {Odelius},\ and\ \citenamefont
  {F\"ohlisch}}]{SUN2011}%
  \BibitemOpen
  \bibfield  {author} {\bibinfo {author} {\bibfnamefont {Y.-P.}\ \bibnamefont
  {Sun}}, \bibinfo {author} {\bibfnamefont {F.}~\bibnamefont {Hennies}},
  \bibinfo {author} {\bibfnamefont {A.}~\bibnamefont {Pietzsch}}, \bibinfo
  {author} {\bibfnamefont {B.}~\bibnamefont {Kennedy}}, \bibinfo {author}
  {\bibfnamefont {T.}~\bibnamefont {Schmitt}}, \bibinfo {author} {\bibfnamefont
  {V.~N.}\ \bibnamefont {Strocov}}, \bibinfo {author} {\bibfnamefont
  {J.}~\bibnamefont {Andersson}}, \bibinfo {author} {\bibfnamefont
  {M.}~\bibnamefont {Berglund}}, \bibinfo {author} {\bibfnamefont {J.-E.}\
  \bibnamefont {Rubensson}}, \bibinfo {author} {\bibfnamefont {K.}~\bibnamefont
  {Aidas}}, \bibinfo {author} {\bibfnamefont {F.}~\bibnamefont {Gel'mukhanov}},
  \bibinfo {author} {\bibfnamefont {M.}~\bibnamefont {Odelius}}, \ and\
  \bibinfo {author} {\bibfnamefont {A.}~\bibnamefont {F\"ohlisch}},\ }\href
  {\doibase 10.1103/PhysRevB.84.132202} {\bibfield  {journal} {\bibinfo
  {journal} {Phys. Rev. B}\ }\textbf {\bibinfo {volume} {84}},\ \bibinfo
  {pages} {132202} (\bibinfo {year} {2011})}\BibitemShut {NoStop}%
\bibitem [{\citenamefont {Hennies}\ \emph {et~al.}(2010)\citenamefont
  {Hennies}, \citenamefont {Pietzsch}, \citenamefont {Berglund}, \citenamefont
  {F\"ohlisch}, \citenamefont {Schmitt}, \citenamefont {Strocov}, \citenamefont
  {Karlsson}, \citenamefont {Andersson},\ and\ \citenamefont
  {Rubensson}}]{HENNIES2010}%
  \BibitemOpen
  \bibfield  {author} {\bibinfo {author} {\bibfnamefont {F.}~\bibnamefont
  {Hennies}}, \bibinfo {author} {\bibfnamefont {A.}~\bibnamefont {Pietzsch}},
  \bibinfo {author} {\bibfnamefont {M.}~\bibnamefont {Berglund}}, \bibinfo
  {author} {\bibfnamefont {A.}~\bibnamefont {F\"ohlisch}}, \bibinfo {author}
  {\bibfnamefont {T.}~\bibnamefont {Schmitt}}, \bibinfo {author} {\bibfnamefont
  {V.}~\bibnamefont {Strocov}}, \bibinfo {author} {\bibfnamefont {H.~O.}\
  \bibnamefont {Karlsson}}, \bibinfo {author} {\bibfnamefont {J.}~\bibnamefont
  {Andersson}}, \ and\ \bibinfo {author} {\bibfnamefont {J.-E.}\ \bibnamefont
  {Rubensson}},\ }\href {\doibase 10.1103/PhysRevLett.104.193002} {\bibfield
  {journal} {\bibinfo  {journal} {Phys. Rev. Lett.}\ }\textbf {\bibinfo
  {volume} {104}},\ \bibinfo {pages} {193002} (\bibinfo {year}
  {2010})}\BibitemShut {NoStop}%
\bibitem [{\citenamefont {Ilakovac}\ \emph {et~al.}(2017)\citenamefont
  {Ilakovac}, \citenamefont {Carniato}, \citenamefont {Foury-Leylekian},
  \citenamefont {Tomi\ifmmode~\acute{c}\else \'{c}\fi{}}, \citenamefont
  {Pouget}, \citenamefont {Lazi\ifmmode~\acute{c}\else \'{c}\fi{}},
  \citenamefont {Joly}, \citenamefont {Miyagawa}, \citenamefont {Kanoda},\ and\
  \citenamefont {Nicolaou}}]{PhysRevB.96.184303}%
  \BibitemOpen
  \bibfield  {author} {\bibinfo {author} {\bibfnamefont {V.}~\bibnamefont
  {Ilakovac}}, \bibinfo {author} {\bibfnamefont {S.}~\bibnamefont {Carniato}},
  \bibinfo {author} {\bibfnamefont {P.}~\bibnamefont {Foury-Leylekian}},
  \bibinfo {author} {\bibfnamefont {S.}~\bibnamefont
  {Tomi\ifmmode~\acute{c}\else \'{c}\fi{}}}, \bibinfo {author} {\bibfnamefont
  {J.-P.}\ \bibnamefont {Pouget}}, \bibinfo {author} {\bibfnamefont
  {P.}~\bibnamefont {Lazi\ifmmode~\acute{c}\else \'{c}\fi{}}}, \bibinfo
  {author} {\bibfnamefont {Y.}~\bibnamefont {Joly}}, \bibinfo {author}
  {\bibfnamefont {K.}~\bibnamefont {Miyagawa}}, \bibinfo {author}
  {\bibfnamefont {K.}~\bibnamefont {Kanoda}}, \ and\ \bibinfo {author}
  {\bibfnamefont {A.}~\bibnamefont {Nicolaou}},\ }\href {\doibase
  10.1103/PhysRevB.96.184303} {\bibfield  {journal} {\bibinfo  {journal} {Phys.
  Rev. B}\ }\textbf {\bibinfo {volume} {96}},\ \bibinfo {pages} {184303}
  (\bibinfo {year} {2017})}\BibitemShut {NoStop}%
\bibitem [{\citenamefont {Monney}\ \emph {et~al.}(2013)\citenamefont {Monney},
  \citenamefont {Bisogni}, \citenamefont {Zhou}, \citenamefont {Kraus},
  \citenamefont {Strocov}, \citenamefont {Behr}, \citenamefont {M\'alek},
  \citenamefont {Kuzian}, \citenamefont {Drechsler}, \citenamefont {Johnston},
  \citenamefont {Revcolevschi}, \citenamefont {B\"uchner}, \citenamefont
  {R\o{}nnow}, \citenamefont {van~den Brink}, \citenamefont {Geck},\ and\
  \citenamefont {Schmitt}}]{MONNEY2013}%
  \BibitemOpen
  \bibfield  {author} {\bibinfo {author} {\bibfnamefont {C.}~\bibnamefont
  {Monney}}, \bibinfo {author} {\bibfnamefont {V.}~\bibnamefont {Bisogni}},
  \bibinfo {author} {\bibfnamefont {K.-J.}\ \bibnamefont {Zhou}}, \bibinfo
  {author} {\bibfnamefont {R.}~\bibnamefont {Kraus}}, \bibinfo {author}
  {\bibfnamefont {V.~N.}\ \bibnamefont {Strocov}}, \bibinfo {author}
  {\bibfnamefont {G.}~\bibnamefont {Behr}}, \bibinfo {author} {\bibfnamefont
  {J.~c.~v.}\ \bibnamefont {M\'alek}}, \bibinfo {author} {\bibfnamefont
  {R.}~\bibnamefont {Kuzian}}, \bibinfo {author} {\bibfnamefont {S.-L.}\
  \bibnamefont {Drechsler}}, \bibinfo {author} {\bibfnamefont {S.}~\bibnamefont
  {Johnston}}, \bibinfo {author} {\bibfnamefont {A.}~\bibnamefont
  {Revcolevschi}}, \bibinfo {author} {\bibfnamefont {B.}~\bibnamefont
  {B\"uchner}}, \bibinfo {author} {\bibfnamefont {H.~M.}\ \bibnamefont
  {R\o{}nnow}}, \bibinfo {author} {\bibfnamefont {J.}~\bibnamefont {van~den
  Brink}}, \bibinfo {author} {\bibfnamefont {J.}~\bibnamefont {Geck}}, \ and\
  \bibinfo {author} {\bibfnamefont {T.}~\bibnamefont {Schmitt}},\ }\href
  {\doibase 10.1103/PhysRevLett.110.087403} {\bibfield  {journal} {\bibinfo
  {journal} {Phys. Rev. Lett.}\ }\textbf {\bibinfo {volume} {110}},\ \bibinfo
  {pages} {087403} (\bibinfo {year} {2013})}\BibitemShut {NoStop}%
\bibitem [{\citenamefont {Ertan}\ \emph {et~al.}(2017)\citenamefont {Ertan},
  \citenamefont {Kimberg}, \citenamefont {Gel'mukhanov}, \citenamefont
  {Hennies}, \citenamefont {Rubensson}, \citenamefont {Schmitt}, \citenamefont
  {Strocov}, \citenamefont {Zhou}, \citenamefont {Iannuzzi}, \citenamefont
  {F\"ohlisch}, \citenamefont {Odelius},\ and\ \citenamefont
  {Pietzsch}}]{ERTAN2017}%
  \BibitemOpen
  \bibfield  {author} {\bibinfo {author} {\bibfnamefont {E.}~\bibnamefont
  {Ertan}}, \bibinfo {author} {\bibfnamefont {V.}~\bibnamefont {Kimberg}},
  \bibinfo {author} {\bibfnamefont {F.}~\bibnamefont {Gel'mukhanov}}, \bibinfo
  {author} {\bibfnamefont {F.}~\bibnamefont {Hennies}}, \bibinfo {author}
  {\bibfnamefont {J.-E.}\ \bibnamefont {Rubensson}}, \bibinfo {author}
  {\bibfnamefont {T.}~\bibnamefont {Schmitt}}, \bibinfo {author} {\bibfnamefont
  {V.~N.}\ \bibnamefont {Strocov}}, \bibinfo {author} {\bibfnamefont
  {K.}~\bibnamefont {Zhou}}, \bibinfo {author} {\bibfnamefont {M.}~\bibnamefont
  {Iannuzzi}}, \bibinfo {author} {\bibfnamefont {A.}~\bibnamefont
  {F\"ohlisch}}, \bibinfo {author} {\bibfnamefont {M.}~\bibnamefont {Odelius}},
  \ and\ \bibinfo {author} {\bibfnamefont {A.}~\bibnamefont {Pietzsch}},\
  }\href {\doibase 10.1103/PhysRevB.95.144301} {\bibfield  {journal} {\bibinfo
  {journal} {Phys. Rev. B}\ }\textbf {\bibinfo {volume} {95}},\ \bibinfo
  {pages} {144301} (\bibinfo {year} {2017})}\BibitemShut {NoStop}%
\bibitem [{\citenamefont {Habicht}\ \emph {et~al.}()\citenamefont {Habicht},
  \citenamefont {Wong},\ and\ \citenamefont {Schulz}}]{DENIZ}%
  \BibitemOpen
  \bibfield  {author} {\bibinfo {author} {\bibfnamefont {K.}~\bibnamefont
  {Habicht}}, \bibinfo {author} {\bibfnamefont {D.}~\bibnamefont {Wong}}, \
  and\ \bibinfo {author} {\bibfnamefont {C.}~\bibnamefont {Schulz}},\
  }\href@noop {} {}\bibinfo {howpublished} {Private communication}\BibitemShut
  {NoStop}%
\bibitem [{\citenamefont {Avella}\ \emph {et~al.}(2018)\citenamefont {Avella},
  \citenamefont {Ole\ifmmode~\acute{s}\else \'{s}\fi{}},\ and\ \citenamefont
  {Horsch}}]{AVELLA2018}%
  \BibitemOpen
  \bibfield  {author} {\bibinfo {author} {\bibfnamefont {A.}~\bibnamefont
  {Avella}}, \bibinfo {author} {\bibfnamefont {A.~M.}\ \bibnamefont
  {Ole\ifmmode~\acute{s}\else \'{s}\fi{}}}, \ and\ \bibinfo {author}
  {\bibfnamefont {P.}~\bibnamefont {Horsch}},\ }\href {\doibase
  10.1103/PhysRevB.97.155104} {\bibfield  {journal} {\bibinfo  {journal} {Phys.
  Rev. B}\ }\textbf {\bibinfo {volume} {97}},\ \bibinfo {pages} {155104}
  (\bibinfo {year} {2018})}\BibitemShut {NoStop}%
\bibitem [{\citenamefont {Miyasaka}\ \emph {et~al.}(2006)\citenamefont
  {Miyasaka}, \citenamefont {Fujioka}, \citenamefont {Iwama}, \citenamefont
  {Okimoto},\ and\ \citenamefont {Tokura}}]{MIYASAKA2006}%
  \BibitemOpen
  \bibfield  {author} {\bibinfo {author} {\bibfnamefont {S.}~\bibnamefont
  {Miyasaka}}, \bibinfo {author} {\bibfnamefont {J.}~\bibnamefont {Fujioka}},
  \bibinfo {author} {\bibfnamefont {M.}~\bibnamefont {Iwama}}, \bibinfo
  {author} {\bibfnamefont {Y.}~\bibnamefont {Okimoto}}, \ and\ \bibinfo
  {author} {\bibfnamefont {Y.}~\bibnamefont {Tokura}},\ }\href {\doibase
  10.1103/PhysRevB.73.224436} {\bibfield  {journal} {\bibinfo  {journal} {Phys.
  Rev. B}\ }\textbf {\bibinfo {volume} {73}},\ \bibinfo {pages} {224436}
  (\bibinfo {year} {2006})}\BibitemShut {NoStop}%
\bibitem [{\citenamefont {Ulrich}\ \emph {et~al.}(2009)\citenamefont {Ulrich},
  \citenamefont {Ament}, \citenamefont {Ghiringhelli}, \citenamefont
  {Braicovich}, \citenamefont {Sala}, \citenamefont {Pezzotta}, \citenamefont
  {Schmitt}, \citenamefont {Khaliullin}, \citenamefont {van~den Brink},
  \citenamefont {Roth}, \citenamefont {Lorenz},\ and\ \citenamefont
  {Keimer}}]{ULRICH2009}%
  \BibitemOpen
  \bibfield  {author} {\bibinfo {author} {\bibfnamefont {C.}~\bibnamefont
  {Ulrich}}, \bibinfo {author} {\bibfnamefont {L.~J.~P.}\ \bibnamefont
  {Ament}}, \bibinfo {author} {\bibfnamefont {G.}~\bibnamefont {Ghiringhelli}},
  \bibinfo {author} {\bibfnamefont {L.}~\bibnamefont {Braicovich}}, \bibinfo
  {author} {\bibfnamefont {M.}~\bibnamefont {Sala}}, \bibinfo {author}
  {\bibfnamefont {N.}~\bibnamefont {Pezzotta}}, \bibinfo {author}
  {\bibfnamefont {T.}~\bibnamefont {Schmitt}}, \bibinfo {author} {\bibfnamefont
  {G.}~\bibnamefont {Khaliullin}}, \bibinfo {author} {\bibfnamefont
  {J.}~\bibnamefont {van~den Brink}}, \bibinfo {author} {\bibfnamefont
  {H.}~\bibnamefont {Roth}}, \bibinfo {author} {\bibfnamefont {T.}~\bibnamefont
  {Lorenz}}, \ and\ \bibinfo {author} {\bibfnamefont {B.}~\bibnamefont
  {Keimer}},\ }\href {\doibase 10.1103/PhysRevLett.103.107205} {\bibfield
  {journal} {\bibinfo  {journal} {Phys. Rev. Lett.}\ }\textbf {\bibinfo
  {volume} {103}},\ \bibinfo {pages} {107205} (\bibinfo {year}
  {2009})}\BibitemShut {NoStop}%
\bibitem [{\citenamefont {Haverkort}\ \emph {et~al.}(2012)\citenamefont
  {Haverkort}, \citenamefont {Zwierzycki},\ and\ \citenamefont
  {Andersen}}]{HAVERKORT2012}%
  \BibitemOpen
  \bibfield  {author} {\bibinfo {author} {\bibfnamefont {M.~W.}\ \bibnamefont
  {Haverkort}}, \bibinfo {author} {\bibfnamefont {M.}~\bibnamefont
  {Zwierzycki}}, \ and\ \bibinfo {author} {\bibfnamefont {O.~K.}\ \bibnamefont
  {Andersen}},\ }\href {\doibase 10.1103/PhysRevB.85.165113} {\bibfield
  {journal} {\bibinfo  {journal} {Phys. Rev. B}\ }\textbf {\bibinfo {volume}
  {85}},\ \bibinfo {pages} {165113} (\bibinfo {year} {2012})}\BibitemShut
  {NoStop}%
\bibitem [{\citenamefont {Haverkort}(2004)}]{HAVERKORT2004}%
  \BibitemOpen
  \bibfield  {author} {\bibinfo {author} {\bibfnamefont {M.}~\bibnamefont
  {Haverkort}},\ }\href@noop {} {Ph.D. thesis},\ \bibinfo  {school} {University
  of Cologne} (\bibinfo {year} {2004})\BibitemShut {NoStop}%
\bibitem [{\citenamefont {Jana}\ \emph {et~al.}(2020)\citenamefont {Jana},
  \citenamefont {Raghunathan}, \citenamefont {Rawat}, \citenamefont
  {Choudhary},\ and\ \citenamefont {Phase}}]{jana2020}%
  \BibitemOpen
  \bibfield  {author} {\bibinfo {author} {\bibfnamefont {A.}~\bibnamefont
  {Jana}}, \bibinfo {author} {\bibfnamefont {R.}~\bibnamefont {Raghunathan}},
  \bibinfo {author} {\bibfnamefont {R.}~\bibnamefont {Rawat}}, \bibinfo
  {author} {\bibfnamefont {R.~J.}\ \bibnamefont {Choudhary}}, \ and\ \bibinfo
  {author} {\bibfnamefont {D.~M.}\ \bibnamefont {Phase}},\ }\href@noop {} {}
  (\bibinfo {year} {2020}),\ \Eprint {http://arxiv.org/abs/2007.03894}
  {arXiv:2007.03894} \BibitemShut {NoStop}%
\bibitem [{\citenamefont {Copie}\ \emph {et~al.}(2013)\citenamefont {Copie},
  \citenamefont {Rotella}, \citenamefont {Boullay}, \citenamefont {Morales},
  \citenamefont {Pautrat}, \citenamefont {Janolin}, \citenamefont {Infante},
  \citenamefont {Pravathana}, \citenamefont {Lüders},\ and\ \citenamefont
  {Prellier}}]{COPIE2013}%
  \BibitemOpen
  \bibfield  {author} {\bibinfo {author} {\bibfnamefont {O.}~\bibnamefont
  {Copie}}, \bibinfo {author} {\bibfnamefont {H.}~\bibnamefont {Rotella}},
  \bibinfo {author} {\bibfnamefont {P.}~\bibnamefont {Boullay}}, \bibinfo
  {author} {\bibfnamefont {M.}~\bibnamefont {Morales}}, \bibinfo {author}
  {\bibfnamefont {A.}~\bibnamefont {Pautrat}}, \bibinfo {author} {\bibfnamefont
  {P.-E.}\ \bibnamefont {Janolin}}, \bibinfo {author} {\bibfnamefont {I.~C.}\
  \bibnamefont {Infante}}, \bibinfo {author} {\bibfnamefont {D.}~\bibnamefont
  {Pravathana}}, \bibinfo {author} {\bibfnamefont {U.}~\bibnamefont {Lüders}},
  \ and\ \bibinfo {author} {\bibfnamefont {W.}~\bibnamefont {Prellier}},\
  }\href {\doibase 10.1088/0953-8984/25/49/492201} {\bibfield  {journal}
  {\bibinfo  {journal} {J. Phys.: Condens. Matter}\ }\textbf {\bibinfo {volume}
  {25}},\ \bibinfo {pages} {492201} (\bibinfo {year} {2013})}\BibitemShut
  {NoStop}%
\bibitem [{\citenamefont {Ishihara}(2004)}]{ISHIHARA2004}%
  \BibitemOpen
  \bibfield  {author} {\bibinfo {author} {\bibfnamefont {S.}~\bibnamefont
  {Ishihara}},\ }\href {\doibase 10.1103/PhysRevB.69.075118} {\bibfield
  {journal} {\bibinfo  {journal} {Phys. Rev. B}\ }\textbf {\bibinfo {volume}
  {69}},\ \bibinfo {pages} {075118} (\bibinfo {year} {2004})}\BibitemShut
  {NoStop}%
\end{thebibliography}%
\end{document}